\documentclass{aa}  

\usepackage{graphicx}
\usepackage{longtable}
\usepackage[utf8]{inputenc} % allow utf-8 input
\usepackage[T1]{fontenc}    % use 8-bit T1 fonts
\usepackage{url}
\usepackage{upgreek}
\usepackage{multirow}

\usepackage[dvipsnames]{xcolor}
\usepackage[colorlinks=true,allcolors=blue]{hyperref}
\usepackage{txfonts}
\usepackage[normalem]{ulem}

\usepackage{txfonts}
\usepackage{color}

\newcommand{\teff}{T$_{\rm eff}$}
\newcommand{\feh}{\ensuremath{\protect\rm [Fe/H] } }

\newcommand{\sect}[1]{\text{Section~\ref{#1}}}
\newcommand{\fig}[1]{\text{Fig.~\ref{#1}}}
\newcommand{\tab}[1]{\text{Table~\ref{#1}}}

\newcommand{\mtd}{\textlangle3D\textrangle}
\newcommand{\multitd}{\texttt{Multi3D}}
\newcommand{\balder}{\texttt{Balder}}
\newcommand{\detail}{\texttt{DETAIL}}

\newcommand{\marcs}{\texttt{MARCS}}

\newcommand{\imbalance}{\Delta_{\textsc{I}-\textsc{II}}}

\begin{document}

   %\title{NLTE titanium abundances in late-type benchmark stars}
    \title{Titanium abundances in late-type stars}
    \subtitle{I. 1D non-LTE modelling in benchmark dwarfs and giants}
    \authorrunning{J. W. E. Mallinson et al.}
    \titlerunning{Titanium abundances in late-type stars}

   %\author{J. Mallinson$^1$, K. Lind$^1$, A. M. Amarsi$^2$, P. Barklem$^2$, J. Grumer$^2$, A. K. Belyaev$^3$}
    \author{J.~W.~E.~Mallinson\inst{\ref{su}}, K.~Lind\inst{\ref{su}},
   A.~M.~Amarsi\inst{\ref{uu1}}, 
   P.~S.~Barklem\inst{\ref{uu1}}, J. Grumer\inst{\ref{uu1}}, 
   A.~K.~Belyaev\inst{\ref{moscow}}, K. Youakim\inst{\ref{su}}}

   %\institute{$^1$ Department of Astronomy, Stockholm University, AlbaNova University Center, SE-106 91 Stockholm, Sweden  \\
   %$^2$ Department of Physics and Astronomy, Uppsala University, Uppsala 75237, Sweden \\  
   %$^3$ Department of Theoretical Physics and Astronomy, Herzen University, St. Petersburg 191186 Russia}
              
    \institute{\label{su}Department of Astronomy, Stockholm
University, AlbaNova University Centre, Roslagstullsbacken, SE-106 91 Stockholm, Sweden\\
\email{jack.mallinson@astro.su.se}
\and
\label{uu1}Theoretical Astrophysics,
Department of Physics and Astronomy,
Uppsala University, Box 516, SE-751 20 Uppsala, Sweden
\and
\label{moscow}Department of Theoretical Physics and Astronomy, Herzen University, St. Petersburg, Riv Moyka, 191186 Russia}

   \date{}

% \abstract{}{}{}{}{} 
% 5 {} token are mandatory
 
  \abstract
   {The titanium abundances of late-type stars are important tracers of Galactic formation history. However, abundances inferred from \ion{Ti}{I} and \ion{Ti}{II} lines can be in stark disagreement in very metal-poor giants. Departures from local thermodynamic equilibrium (LTE) have a large impact on the minority neutral species and thus influence the ionisation imbalance, but satisfactory non-LTE modelling for both dwarfs and giants has not been achieved in the literature.}
   {The reliability of titanium abundances is reassessed in benchmark dwarfs and giants using a new non-LTE model 1D model atmospheres.}
   {A comprehensive model atom was compiled with a more extended level structure and newly published data for inelastic collisions between \ion{Ti}{I} and neutral hydrogen.}
   {In 1D LTE, the \ion{Ti}{I} and \ion{Ti}{II} lines agree to within 0.06 dex for the Sun, Arcturus, and the very metal-poor stars HD84937 and HD140283.  For the very metal-poor giant HD122563, the \ion{Ti}{I} lines give an abundance that is 0.47 dex lower than that from \ion{Ti}{II}. 
   The 1D non-LTE corrections can reach $+0.4$ dex for individual \ion{Ti}{I} lines and $+0.1$ dex for individual \ion{Ti}{II} lines, and they reduce the overall ionisation imbalance to $-0.17$ dex for HD122563.  However, the corrections also increase the imbalance for the very metal-poor dwarf and sub-giant to around 0.2 dex.}
  % conclusions heading (optional), leave it empty if necessary https://www.overleaf.com/project/60dafb40afb37556496925f8
   {Using 1D non-LTE reduces the ionisation imbalance in very metal-poor giants but breaks the balance of other very metal-poor stars, consistent with conclusions drawn in the literature. To make further progress, consistent 3D non-LTE models are needed.} %To make further progress, the impact on a larger sample of stars, as well as the potential of 3D non-LTE effects, should be investigated.}
   
   \titlerunning{Titanium abundances in late-type stars}
   \maketitle
%
%-------------------------------------------------------------------

\section{Introduction}
\label{introduction}    
    
    In the era of the European Space Agency \textit{Gaia} mission \citep{Gaia2016} and large spectroscopic datasets from current and upcoming million-star surveys such as the Large Sky Area Multi-Object Fibre Spectroscopic Telescope \citep{Zhao12}, Apache Point Observatory Galactic Evolution Experiment \citep{Majewski17}, William Herschel Telescope Enhanced Area Velocity Explorer \citep{Dalton18}, 4-metre Multi-Object Spectrograph Telescope \citep{deJong19}, and the Galactic Archaeology with High-Efficiency and high-Resolution Mercator Echelle Spectrograph \citep{Buder21}, precision spectroscopy is becoming increasingly important for understanding the history and evolution of the Milky Way thanks to the accuracy and detail of the information being received. For example, current and future observations of stars with peculiar chemical abundance patterns, especially in the metal-poor regime, shed light on the assembly of the early Galaxy and its enrichment by supernova explosions \citep[][]{Nissen18, Helmi20}.

    In this context, titanium is an element of high astrophysical interest. \ion{Ti}{I} and \ion{Ti}{II} lines are observed throughout a wide range of stars and produce numerous spectral lines. Thus, they are commonly used to calculate titanium abundances and fundamental stellar parameters such as effective temperature (\teff{}) and surface gravity ($\log (g)$), and can also be a proxy for metallicity. Moreover, as titanium is an $\upalpha$-element, its abundances can trace stars of different ages \citep{stellarage} due to the titanium output difference between Type Ia and core-collapse supernovae. Hence, the Galactic thin and thick discs separate out in the [Ti/Fe] versus [Fe/H] plane \citep{disks}. As such, accurate titanium abundance measurements can give insight into the formation and evolution of the Galaxy.
    
    The success of these types of studies critically depends on the accuracy of the inferred titanium abundances. In late-type stars, it is usually possible to infer titanium abundances from both \ion{Ti}{I} and \ion{Ti}{II} lines.  The level of ionisation balance, $\imbalance\equiv\mathrm{A\left(Ti\right)_{\ion{Ti}{I}}-A\left(Ti\right)_{\ion{Ti}{II}}}$, can therefore be measured, with non-zero values indicative of some deficiency in the spectral models.  
    
    One such deficiency could be the assumption of local thermodynamic equilibrium (LTE), as commonly assumed in
    classical spectroscopic analyses. In this approximation, the populations of excited and ionised states of titanium follow the Saha-Boltzmann distributions.  In reality, similarly to neutral iron, the neutral minority species is prone to departures from LTE: the supra-thermal ultraviolet (ultraviolet) radiation field typically leads to over-ionisation \citep{berge}. This effect grows towards lower metallicities, where ultraviolet photons can travel through the atmosphere with even less impediment; thus, both the excess radiation and the LTE error increase.
    
    For example, \citet{scott} find $\imbalance=-0.15$ dex for the Sun when using a 3D hydrodynamic model solar atmosphere and 3D LTE radiative transfer. By employing non-LTE corrections computed on a \mtd{} model atmosphere, \citet{scott} find a better agreement between the two species, $\imbalance=-0.09$ dex. As the authors discuss, consistent 3D non-LTE modelling using improved data for inelastic collisions with neutral hydrogen may be needed to fully resolve these remaining ionisation imbalances.
    
    For stars other than the Sun, all titanium abundance analyses to date have been based on 1D model atmospheres.  In the absence of 3D non-LTE models, 1D non-LTE is expected to be more reliable than 1D LTE or 3D LTE, at least for neutral iron \citep{anishFe,nordlander}. The most comprehensive 1D non-LTE models to date were recently presented by \citet{Sitnova_2020}.  The authors used the 1D non-LTE code \detail{} \citep{detailcode} and employed a large model atom utilising, for the first time, ab initio inelastic hydrogen collisions. For the Sun, they find $\imbalance=-0.07$ in 1D LTE, which is in fact similar to the 1D LTE results presented by \citet{scott}. \citet{Sitnova_2020} find this improves to $-0.03$ dex in 1D non-LTE.  
    
    However, there are larger discrepancies in the metal-poor regime.  For the very metal-poor giant HD122563, \citet{Sitnova_2020} report $\imbalance=-0.4$ in 1D LTE.  This becomes less severe in 1D non-LTE, $\imbalance=-0.2$.  Unfortunately, 1D non-LTE instead worsens the ionisation imbalance for the other very metal-poor stars. In 1D LTE, they find $\imbalance=+0.05$ and $-0.03$ dex for the benchmark stars HD84937 and HD140283; these change to $0.17$ and $0.11$ dex, respectively.

    It is not clear from where exactly the large ionisation imbalances found by \citet{Sitnova_2020} in 1D non-LTE originate.  In this context, it is important to note that different groups have reported different 1D non-LTE results for other iron-peak elements. In particular, for copper, \citet{shiCu} report discrepancies of $0.25$ dex, $0.38$ dex, and $0.69$ dex for HD84937, HD140283, and HD122563, respectively, between their 1D non-LTE results and those of \citet{andrievskyCu}.  As such, it is worthwhile to test whether the ionisation imbalances can be due to deficiencies in the non-LTE models, rather than, for example, failures of the 1D model atmospheres. 
    
    This work presents an independent 1D non-LTE study of titanium abundances in late-type benchmark stars. Compared to \citet{Sitnova_2020}, the results presented here are based on a different non-LTE code and a new model atom with higher un-collapsed energy levels utilising a more up-to-date prescription for the inelastic collisions with neutral hydrogen. This non-LTE model is used in an attempt to solve the imbalance for the Sun, dwarfs, and giants at once.
    The rest of this article is structured as follows.
    The non-LTE model is presented in \sect{method},
    and the analysis is described in \sect{analysis}.
    The results for the Sun, the giant Arcturus, 
    the very metal-poor stars HD84937 and HD140283,
    and the very metal-poor giant HD122563 are presented in \sect{results}, and the work is concluded in \sect{conclusions}.

    %Due to its size and complexity, as well as the atmosphere simulations used, it is still being run in 1D atmospheres.

%--------------------------------------------------------------------
\section{Method}
\label{method}

\subsection{Overview}
\label{method_model}

% KL: This information is not appropriate here. Needs to move later into the text and described 
%The model will still use assumptions of hydrogenic approximations \citep{hydrogenic} for a minority of photo-ionisations, the \citet{kaulakys} formula for the less important hydrogenic collisions that occur above the ionic limit, and \citet{regemorter} approximations for electron collisional data. Accurate radiative transitions for the majority of over 50,000 energy levels from the \citet{K16} titanium atom were collected. This solves many of the issues mentioned in previous papers such as lack of high-excitation levels, Drawin approximations, and too many hydrogenic photo-ionisation cross section. \citep{berge}. The hydrogen collisions used are also more accurate than previously used work \citep{Sitnova_2020} and so produce a different result. Furthermore, many metal-poor stars that failed to achieve a small ionisation imbalance were analysed to show that the model atom works in a wide range of stars.

The non-LTE calculations and theoretical stellar spectra were performed using \balder{} \citep{balder}.  This code is based on \multitd{} \citep{leenaart} with updates to the parallelisation scheme and background opacities \citep{balder} and the statistical equilibrium solver \citep{anishC}.
%The reduced model atom is available at XXXX.com as a .fits file which consists of four extensions containing the energy levels and transition cross-sections.

\begin{figure*}
\centering
\includegraphics[width=15cm]{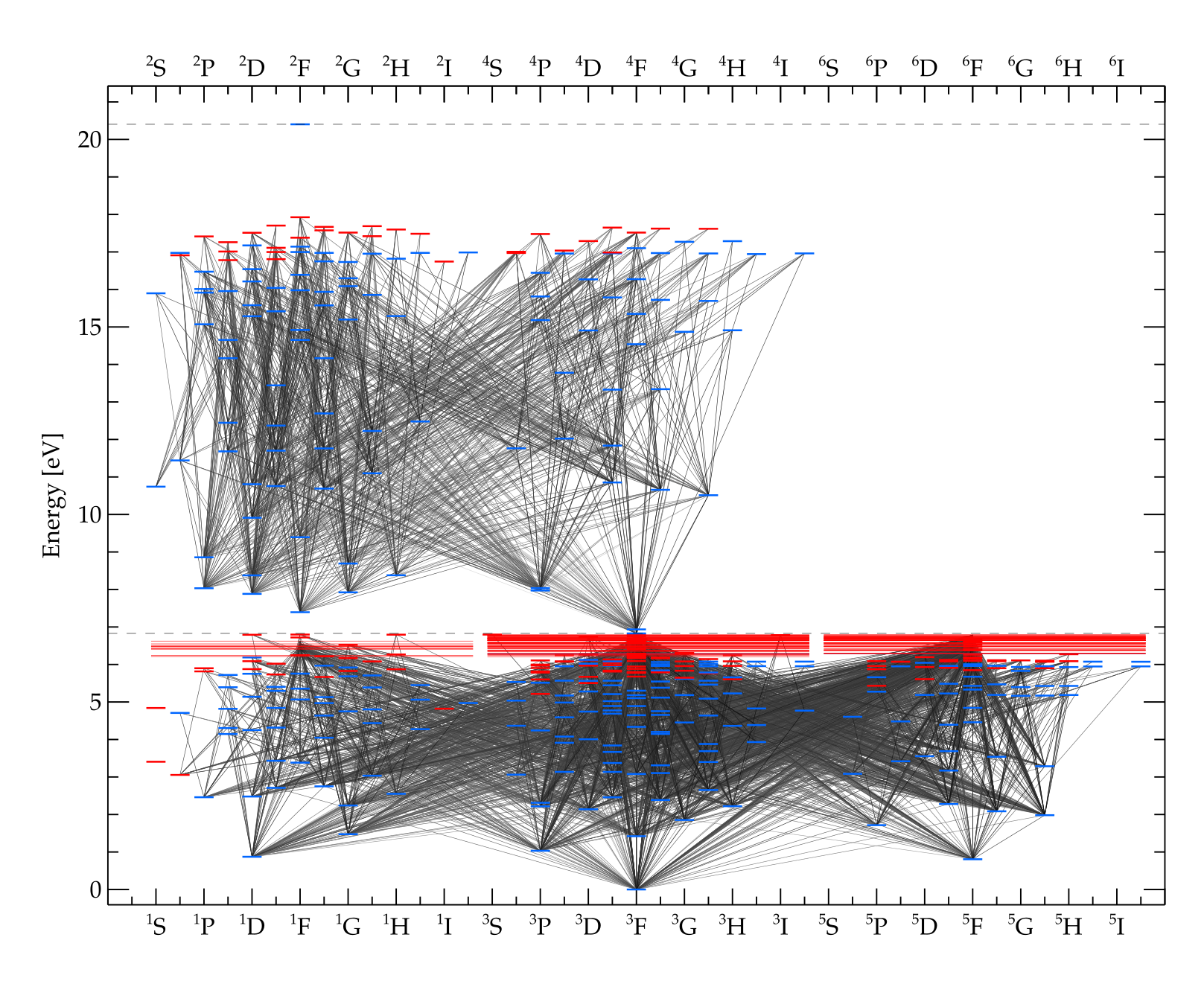}
  \caption{Term diagram of the reduced atom showing the energies for possible bound-bound transitions of \ion{Ti}{I} (lower half) and \ion{Ti}{II} (upper half). Transitions are shown by black lines, with darker lines representing a higher oscillator strength. Blue marks show the observed energy levels, and red marks show levels that are theoretically predicted.}
     \label{term}
\end{figure*}

Sections \ref{method_energies}--\ref{electron_collisions} describe the raw data based on which the model atom was constructed. \sect{atom_reduction} then describes the reduction process of the atom.  In summary, the reduced model atom, illustrated in \fig{term}, contains the following:

587 energy levels, of which 459 are of \ion{Ti}{I} and 127 are of \ion{Ti}{II}, and the ground state of \ion{Ti}{III} is included; $4\,784$ bound-bound radiative transitions; and 586 photo-ionisation transitions.% with 3,431,765 total cross sections.

\subsection{Energy levels}
\label{method_energies}

The predicted and observed energy levels of \ion{Ti}{I} and \ion{Ti}{II} were taken from \citet{K16}\footnote{\url{kurucz.harvard.edu/atoms/2200}}\,\footnote{\url{kurucz.harvard.edu/atoms/2201}}, which was also used by \citet{Sitnova_2020}, and are included in the Vienna Atomic Line Database 3 (VALD3) \citep{vald} database. Specifically, the files called gf220X.gam and gf220X.lin were used, where X is 0 or 1. 
The \citet{K16} database contains nearly $18\,000$ bound fine-structure levels for \ion{Ti}{I} and \ion{Ti}{II}.
These level data were combined and reduced as discussed in \sect{atom_reduction}.

\subsection{Radiative transitions}

Data for around 6 million bound-bound transitions were extracted from \citet{K16}. This dataset contains more than 5 million transitions for \ion{Ti}{I} alone. These data were reduced as described in \sect{atom_reduction}. For a subset of lines, experimentally measured f-values from \citet{HD84937} and \citet{HD84937-2} were used. These are the same lines that are used for the abundance analysis in \sect{analysis}.

Photo-ionisation cross-sections for \ion{Ti}{I} were adopted from the \citet{nahardatabse} database\footnote{\url{https://norad.astronomy.osu.edu/}}. Levels from \citet{K16} were cross-referenced to match the Nahar-OSU-Radiative-Atomic-Data (NORAD) database for the initial and target states of photo-ionisation transitions from \citet{nahardatabse} by comparing their electron configurations and terms. Unique matches were found for all but 167 highly excited \ion{Ti}{I} levels in the reduced model atom, none of which had an energy below 3.7eV; the majority were above 6.2eV. For these unmatched levels, the hydrogenic approximation was used \citep{hydrogenic}.

\subsection{Inelastic hydrogen collisions}
\label{hydrogencolls}
Inelastic hydrogen collision data are needed for both bound-bound and charge transfer processes. They are particularly important for metal-poor stellar atmospheres due to the lower electron densities. Therefore, changes in hydrogen collision rate coefficients may have a large impact on the spectra produced. The impact of hydrogen collision processes on the spectra and thus the abundances derived has been shown to be important for other elements \citep{bergemann_Mn, lindbarklem2009, osorio2015, anishOH, reggiani2019, anishC, sitnova_Zn}. 

Initial studies of the non-LTE effects on titanium in stellar atmospheres \citep{berge, sitnova2016} were carried out using the Drawin formula \citep{drawin1968formelmassigen, drawin1969, steenbock1984statistical,lambert_hydrogen}, allowing a scaling factor $S_{\rm H}$ to be chosen to minimise the scatter in derived abundances across observed lines in the Sun. These works found that describing the hydrogen collisions in this manner did not work well for metal-poor stars ($\feh < -2$), giving an ionisation imbalance between \ion{Ti}{I} and \ion{Ti}{II} that was larger than that found later using quantum mechanical calculations for hydrogen collisions \citep{Sitnova_2020}. This is not unexpected, due to the large discrepancy between the Drawin formula results and those from full quantum mechanical calculations \citep{drawin}.

Unfortunately, full quantum mechanical calculations of such processes are to date only available for neutral lithium \citep{bely2003,Barklem2003InelasticHA}, sodium \citep{bely1999,barklemsodium,bely2010}, magnesium \citep{AndreyMg, barklemmagnesium, chemicalMg}, and calcium \citep{andreyCa}.  For other species, especially more complex atoms, one must resort to asymptotic model calculations based on simplified electronic structure and collision dynamics. Such approaches include the asymptotic model from \citet{bely2013}, the asymptotic model from \citet{barklem2016b}, and the simplified method of \citet{belyetal2017SimplifiedModel} and \citet{belyVoronov2018Simplifiedmodel}. The asymptotic model of \citet{barklem2016b} is based on linear combinations of atomic orbitals (LCAO) for the ionic-covalent interactions at avoided ionic crossings, which are expected to be the dominant mechanism where applicable. The others are based on a fit to calculations where the ionic-covalent interactions are taken from a semi-empirical formula \citep{olson1971estimation}.

\begin{figure}
\centering
\includegraphics[width=\hsize]{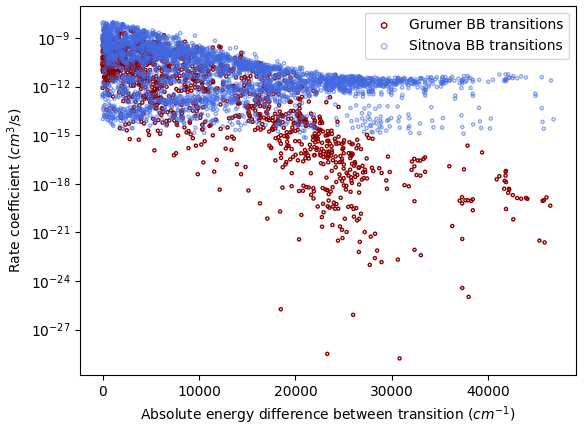}
  \caption{\ion{Ti}{I} bound-bound de-excitation rate coefficients as a function of the energy difference between transitions. Red points are from \citet{grumer}, and blue points are from \citet{Sitnova_2020}.}
     \label{gvb}
\end{figure}

\begin{figure}
\centering
\includegraphics[width=\hsize]{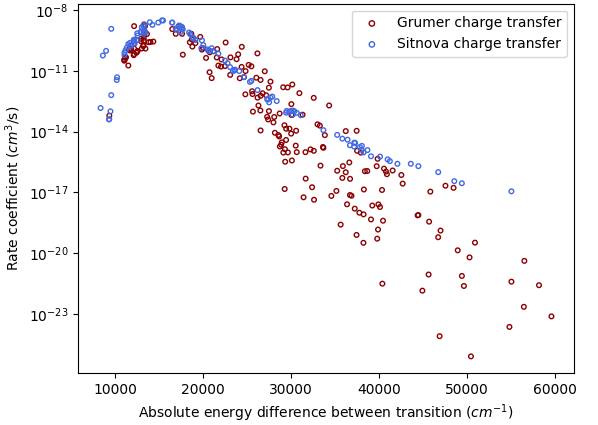}
  \caption{Comparison of charge transfer rates for the reaction $\rm \ion{Ti}{I} + H \rightarrow \ion{Ti}{II} + H^-$. Red points show data from \citet{grumer} and blue points data from \citet{Sitnova_2020}, as in Fig. \ref{gvb}.}
     \label{ctcomparison}
\end{figure}

Asymptotic model calculations based on LCAO were carried out by \citet{grumer} for processes on neutral titanium, while the simplified method was applied in \citet{Sitnova_2020} to both neutral and singly ionised titanium. 
%The simplified \citet{Sitnova_2020} model is tailored towards high and moderate rate coefficients for inelastic low-energy hydrogen collisions of titanium. In contrast, the method of \citet{grumer} allows a better focus on the lower collisional rates of \ion{Ti}{I} but does not include much data on \ion{Ti}{II} levels and transitions. 
The bound-bound excitation rates for neutral titanium are compared in \fig{gvb}, and the charge transfer (ion-pair production) rates involving neutral titanium in \fig{ctcomparison}.  In \fig{ctcomparison} good agreement can be seen for the most important charge transfer transitions: those with large rate coefficients. However, the two datasets differ for processes with low rates, a behaviour also seen for bound-bound processes in \fig{gvb}. The larger scatter seen in the data from \citet{grumer} is due to the explicit consideration of angular momentum coupling in the LCAO approach (that is, the $L$ and $S$ quantum numbers in $LS$ coupling), which is not treated in the simplified approach.

\begin{figure}
\centering
\includegraphics[width=\hsize]{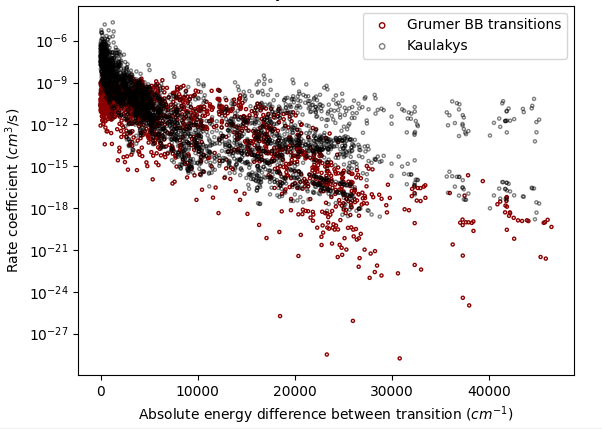}
  \caption{Comparison between de-excitation rate coefficients from \citet{grumer} (red points) and the rates calculated using the Kaulakys code for the same transitions (black points). The relationship found here is the same as in \citet{anishOH}.}
     \label{kaulakys_graph}
\end{figure}

In this work, the collision rate coefficients for bound-bound transitions in \ion{Ti}{I} and its charge transfer rates to \ion{Ti}{II} were taken from \citet{grumer}\footnote{https://github.com/barklem/public-data}, and \ion{Ti}{II} bound-bound rates were taken from \citet{Sitnova_2020}\footnote{http://www.non-lte.com/ti\_h.html}. This resulted in data for the most important transitions, but not all.  For bound-bound transitions in \ion{Ti}{I} involving states above the ionic limit, where the energy is within 0.754eV (the electron affinity of hydrogen) of the ionised ground state, the ionic crossing mechanism does not apply, and an alternate mechanism must be at work.  In these cases, the \citet{kaulakys} free electron model can be used to estimate the cross-sections and rate coefficients, via a momentum transfer mechanism.  It has been argued in earlier work \citep{anishOH} that the contribution from this mechanism should be added to the ionic crossing contribution, though the validity of the free electron model for low-lying states is questionable.  The results of the free electron model calculations for all bound-bound transitions are shown in \fig{kaulakys_graph} and compared to the \citet{grumer} results. Following \citet{anishOH}, the two contributions were added together and their impact on abundance is found to be below 0.1 dex in all stars.

\subsection{Electron collisions}
\label{electron_collisions}
Titanium has no advanced quantum mechanical data for electron collisions for a majority of states, and so all transitions were calculated using the semi-empirical recipes from \citet{regemorter} for excitation and \citet{allen} for ionisation. Due to the dominance of hydrogen collisions, especially in metal-poor stars, the approximations made for electrons are found to have a small impact on the resulting abundances. 

\subsection{Atom reduction}
\label{atom_reduction}
The complete model atom contains thousands of levels and millions of radiative transitions, and it required reduction to make the non-LTE calculations feasible, even in 1D. To begin with, all fine structure was removed. Moreover, \ion{Ti}{I} levels above $6.2$eV with identical parity, electron configuration, and multiplicity were merged to form super-levels and affected lines merged to form super-lines, following \citet{lind17}. 
This created a smaller atom of 587 total energy levels and $50\,000$ transitions, which was still too large to be practical. \ion{Ti}{II} levels within 2.7eV of the ionisation limit were cut due to their low population in late-type stellar atmospheres.

Further reductions were made by considering the radiative brackets $|n_i\,R_{ij} - n_j\,R_{ji}|$ for different bound-bound radiative transitions. This quantity gives an estimate of the relative importance of a radiative transition to the statistical equilibrium: the lower the value, the less important it is and the safer it is to remove it. While this requires an individual depth point to be chosen, the choice does not exert much influence on the overall transition hierarchy \citep{lind17}. By comparing the radiative bracket of the Sun and HD84937, $4\,784$ of the most important bound-bound radiative transitions were selected, and the rest were discarded, resulting in the final reduced atom.

To reduce the computational cost further, the photo-ionisation cross-sections were interpolated onto a common wavelength grid with fixed logarithmic steps, reducing the number of unique wavelength values from over 3 million to just over 38\,000. This reduced computation time and memory requirements since, although the total number of wavelengths did not vary dramatically, it allowed \balder{} to treat each set of identical wavelength values as a single one during the run. While unlikely to cause a significant change, the accuracy of the interpolation was checked in two ways: first, by removing 10\% of the \citet{nahar2015} data before interpolation, re-calculating their values, and comparing them to the original cross-sections of the removed data; secondly, by manual inspection when interpolating all wavelengths to ensure all resonances and features were still retained in the interpolated model.

The final reduced atom is illustrated in \fig{term}, which demonstrates the complexity of the atom, even after collapsing levels.  It contains 459 \ion{Ti}{I} levels under its ionisation potential of $6.828$eV, 127 \ion{Ti}{II} levels that reach up to $2.5$eV below the ionisation limit, and the \ion{Ti}{III} ground state. This atom was run in non-LTE to produce the departure coefficients, $n_\mathrm{\rm NLTE}/n_\mathrm{\rm LTE}$, shown in \fig{sundeparture}. When generating synthetic spectra for diagnostic spectral lines, the departure coefficients were redistributed onto the complete model atom, which contains fine structure, but with the theoretical levels removed.

\subsection{Atom comparison}

This atom contains more unmerged levels than used previously in \citet{Sitnova_2020}. This was done to examine the influence these high energy levels have on titanium abundance predictions. More levels are also coupled with hydrogen collisions via the use of the Kaulakys principle for the higher energy levels above the ionic limit of \ion{Ti}{I}. The average $f\rm{-values}$ are calculated in this paper using the experimental data of \citet{HD84937} and \citet{HD84937-2} where possible, whereas \citet{Sitnova_2020} used the database of R. Kurucz, although it was stated that the comparison was made, and the shift found to be minor, at an average of $\rm{log}(gf_{\rm{lab}})-\rm{log}(gf_{\rm{Kurucz}}) = -0.05\pm0.28$ dex.

\section{Analysis of benchmark stars}
\label{analysis}
Five well-known benchmark stars were analysed using high-resolution optical spectra for the Sun, Arcturus, HD84937, HD140283, and HD122563. The same observational data used in \citet{scott} were used for the Sun, and the observational data from \citet{KarinParameters} were used for all other stars. The same stellar parameters were adopted as in that work, shown in \tab{tablestars}.
In summary, \citet{KarinParameters} used interferometric \teff{} for Arcturus, HD122563, and HD140283, while \teff{} for HD84937 was calculated from its surface-brightness relationship \citep{startableb}. Except for the Sun, $\log (g)$ was computed from mass to radius relations for all stars. \feh{} and microturbulence parameters were adopted from  \citet{KarinParameters}, who determined them simultaneously by enforcing a flat trend in LTE abundance and an equivalent width of \ion{Fe}{II} lines. 
The atmospheres were tailored for each star by interpolating \citep{MasseronPHD} a grid standard of \marcs{} models \citep{MARCSdatabase} onto these stellar parameters.

\begin{table}
\begin{center}
\caption{Stellar parameters of the studied benchmark stars.}
\label{tablestars}
\begin{tabular}{l r r r r r}
\hline
\hline
\noalign{\smallskip}
Star & $T_{\mathrm{eff}}$ & $\log(g)$ & [Fe/H] & $\xi$ & Ref. \\
 & K & $\mathrm{cm\,s^{-2}}$ & & $\mathrm{km\,s^{-1}}$ & \\
\noalign{\smallskip}
\hline
\hline
\noalign{\smallskip}
     Sun & 5772 & 4.44 & $0.00$ & 0.9 & a, d \\
     Arcturus &4286 & 1.64 &$-0.53$ &1.3 & b, d \\
     HD84937 & 6356 & 4.06 &$-2.06$ &1.2 & b, d \\
     HD140283 &5792 & 3.65 &$-2.36$ &1.3 & c, d \\
     HD122563 &4636 &1.40 &$-2.50$ & 1.8 & c, d \\
\noalign{\smallskip}
\hline
\hline
\end{tabular}
\end{center}
\tablefoot{\tablefoottext{a}{Solar \teff{} and $\log (g)$ from \citet{startablea}.} 
\tablefoottext{b}{\teff{} and $\log (g)$ from \citet{startableb}.} 
\tablefoottext{c}{\teff{} and $\log (g)$ from \citet{startablec}.} 
\tablefoottext{d}{\feh{} and $\xi$ from \citet{KarinParameters}.}}
\end{table}

The line selection of titanium can play a large role in some abundance calculations due to uncertain oscillator strengths and blends, as well as difficulties associated with core saturation. Measures were taken to reduce their impact on the overall estimates of titanium abundances, and the final choices are shown in Table \ref{appxtable}.  The impact of uncertain oscillator strengths was mitigated by using a large selection of lines, assuming the uncertainties to be normally distributed.  Spectral lines were fit with one or several Gaussian profiles following \citet{KarinParameters}, from which their equivalent widths were determined, with blended lines removed essentially via sigma-clipping. Saturated lines were removed by imposing a limit on the reduced equivalent width of $\rm{W_{\lambda, red} = log_{10}(W_{\lambda}/\lambda)} = -4.9$.  The final line selection closely resembles that of \citet{scott} for the Sun, \citet{startableb} for Arcturus, and \citet{HD84937} and \citet{HD84937-2} for the very metal-poor stars. 13-37 \ion{Ti}{I} lines and 3-92 \ion{Ti}{II} lines were used, depending on the star.

\section{Results}
\label{results}
\begin{table*}
\begin{center}
\caption{Titanium abundances, A(Ti), and ionisation imbalances, $\imbalance$,  found in this work as well as the unweighted values from \citet{arcturus} for Arcturus, and from \citet{Sitnova_2020} for the other stars.}
\label{tablestarresults}
\begin{tabular}{l l | r r r |  r r r}
\hline
\hline
\noalign{\smallskip}
\multirow{2}{*}{Star} & \multirow{2}{*}{Species} & 
\multicolumn{3}{c|}{LTE} & \multicolumn{3}{|c}{Non-LTE} \\ 
& & Here & Lit. & Diff. & Here & Lit. & Diff.  \\ 
\noalign{\smallskip}
\hline
\hline
\noalign{\smallskip}
     \multirow{3}{*}{Sun} & 
        \ion{Ti}{I}  & 4.87 & 4.88 &  $-0.01$ & 4.90 & 4.91 & $-0.01$ \\
      & \ion{Ti}{II} & 4.92 & 4.95 &  $-0.03$ & 4.92 & 4.94 & $-0.02$ \\
      & $\imbalance$   & $-0.06$   & $-0.07$  & $0.02$ 
                       & $-0.02$   & $-0.03$  &  0.02\\
\noalign{\smallskip}
\noalign{\smallskip}
    \multirow{3}{*}{Arcturus} & 
        \ion{Ti}{I}  & 4.85 & 4.66 &  $0.19$ & 4.86 & ---- & ---- \\
      & \ion{Ti}{II} & 4.93 & 4.66 &  $0.27$ & 4.91 & ---- & ---- \\
      & $\imbalance$   &$-0.08$    & $0.00$ &  $-0.08$ &
                        $-0.05$    & ---- &  ----  \\

\noalign{\smallskip}
\noalign{\smallskip}
    \multirow{3}{*}{HD84937} & 
        \ion{Ti}{I}  & 3.20 & 3.19 &  $0.01$ & 3.40 & 3.35 & $0.05$ \\
      & \ion{Ti}{II} & 3.20 & 3.14 &  $0.06$ & 3.21 & 3.18 & $0.03$ \\
      & $\imbalance$   & $0.00$    & $0.05$ &  $-0.05$ &
                         $0.18$    & $0.17$ &  $-0.01$  \\

\noalign{\smallskip}
\noalign{\smallskip}
    \multirow{3}{*}{HD140283} & 
        \ion{Ti}{I}  & 2.69 & 2.66 &  $0.03$ & 2.91 & 2.84 & $0.07$ \\
      & \ion{Ti}{II} & 2.69 & 2.69 &  $0.00$ & 2.71 & 2.73 & $-0.02$ \\
      & $\imbalance$   &$0.00$     & $-0.03$ &  $-0.03$ &
                        $0.20$     & $0.11$ &  $0.09$  \\

\noalign{\smallskip}
\noalign{\smallskip}
    \multirow{3}{*}{HD122563} & 
        \ion{Ti}{I}  & 2.09 & 2.19 &  $-0.10$ & 2.41 & 2.40 & $0.01$ \\
      & \ion{Ti}{II} & 2.54 & 2.58 &  $-0.04$ & 2.57 & 2.60 & $-0.03$ \\
      & $\imbalance$   & $-0.45$   & $-0.40$ &  $-0.05$  &
                         $-0.16$   & $-0.20$ &  $0.04$ \\

\noalign{\smallskip}
\hline
\hline
\end{tabular}
\end{center}
%\tablebib{(a)~\citet{Sitnova_2020};
%(b)~\citet{arcturus};
%(c)~\citet{HD84937} and \citet{HD84937-2};
%(d)~\citet{hd14};
%(e)~\citet{anotherhd12}}
\end{table*}
\subsection{Overview}

\begin{figure}
\centering
\includegraphics[width=\hsize]{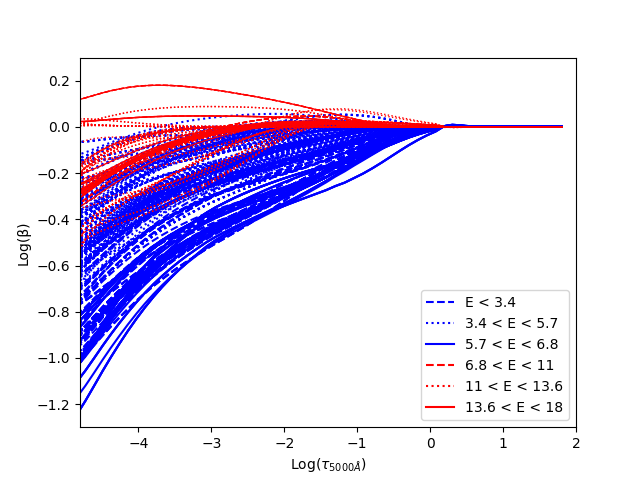}
  \caption{Departure coefficients of titanium in HD84937 as a function of optical depth. Dotted lines represent \ion{Ti}{II} and solid lines represent \ion{Ti}{I}. They reach a rough agreement with LTE deeper in the star as collisions dominate. \ion{Ti}{II} can be seen to be in closer agreement with LTE than \ion{Ti}{I} over all energy levels.}
     \label{sundeparture}
\end{figure}

As anticipated, non-LTE is found to, in general, reduce the strengths of \ion{Ti}{I} lines, consistent with over-ionisation. The departure coefficients for the levels lie below 0 (\fig{sundeparture}).  This translates to positive non-LTE abundance corrections for this species, increasing the titanium abundances inferred from these lines. Compared to \citet{Sitnova_2020}, \fig{sundeparture} misses several \ion{Ti}{II} lines that extend to large departure coefficients. Indeed, this work finds smaller non-LTE effects for \ion{Ti}{II} as this would indicate, as it shows a closer coupling to the LTE value of $\rm{log}_{10}(\beta) = 0$.

The non-LTE corrections are found to be largest for the very metal-poor giant HD122563, where they can reach over +0.4 dex for certain \ion{Ti}{I} lines.  For the very metal-poor dwarf and sub-giant, the non-LTE corrections are of the order +0.2 dex.  For the Sun, the corrections are more muted, only of the order +0.05 dex. Nevertheless, they have an impact on the ionisation balance (\sect{results_sun}).  For Arcturus, the abundance corrections are typically of the order +0.01 dex. 

For \ion{Ti}{II}, the non-LTE corrections are smaller. They can be positive or negative, depending on the transition and on the star. It can be seen that the departure coefficients, given the deeper creation point and majority status of\ion{Ti}{II}, exist closer to LTE in almost all cases. The most severe case is HD122563, where certain lines are affected to the +0.2 dex level; this is much lower in all other stars.

In many stars, but most particularly Arcturus, over-saturation was found when analysing the reduced equivalent widths of the lines and their abundance predictions. To counter the inaccurate abundance estimations due to the modified relationship with abundance the saturated lines have, those with a reduced equivalent width of $\mathrm{log}_{10}(W_{\lambda}/\lambda > -4.9)$ were removed, improving the imbalance in all stars but the Sun in LTE. In Arcturus, this left only three \ion{Ti}{II} lines, but achieved a better ionisation imbalance nonetheless.

\subsection{Abundances and ionisation balance}

\begin{figure*}
\centering
\includegraphics[width=17cm,scale=1.7]{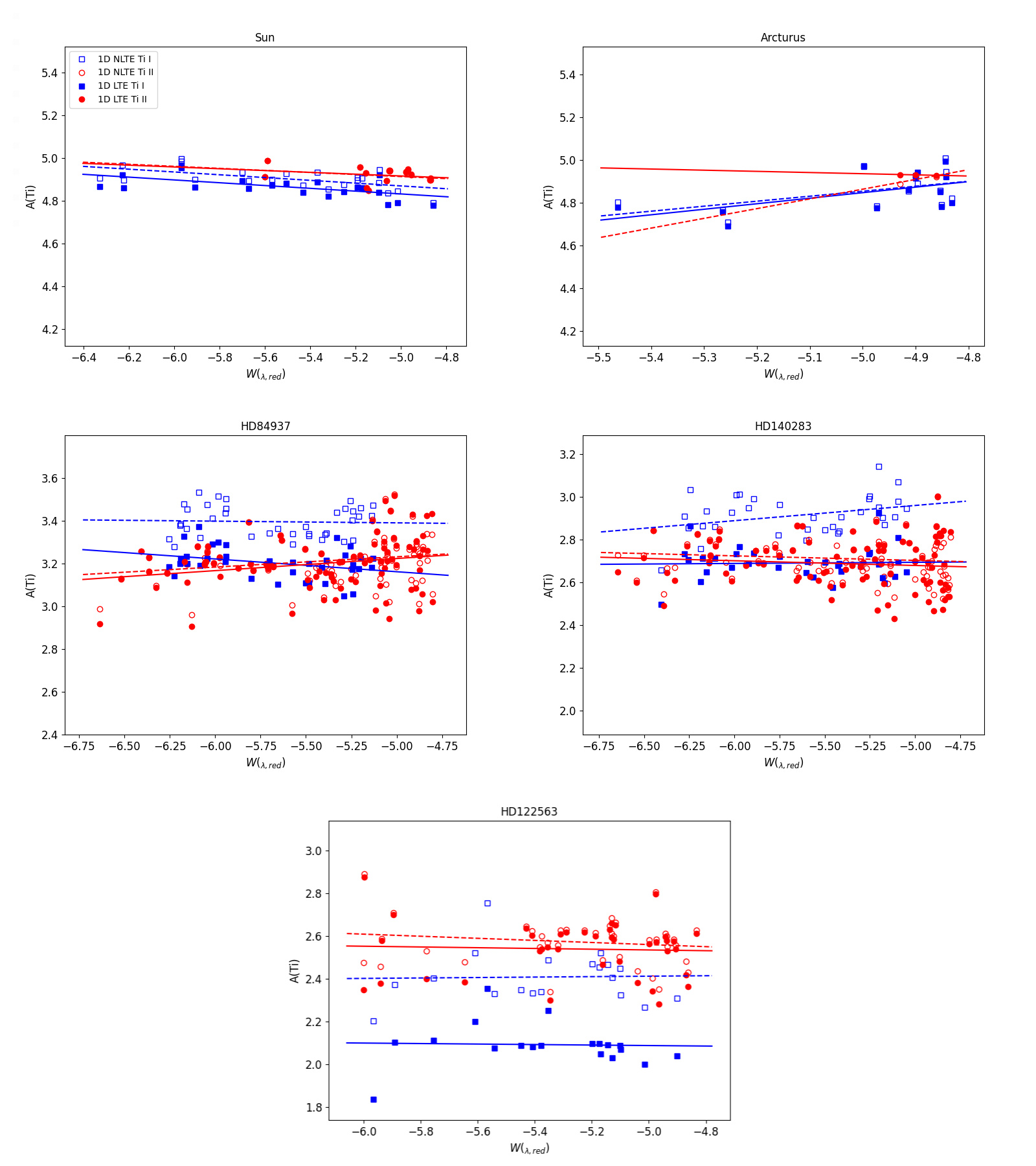}
  \caption{Difference between the mean abundance of \ion{Ti}{I} and \ion{Ti}{II} in LTE and non-LTE for all stars, representing the ionisation imbalance. Filled symbols represent LTE and empty symbols non-LTE. Red circles and lines show \ion{Ti}{II} and blue squares \ion{Ti}{I}. Lines represent the abundance trend, dashed for non-LTE and solid for LTE. The x-axis is the reduced equivalent width: ${\rm log_{10}}(W_{\rm \lambda}/{\rm \lambda}$).}
     \label{ionimbalanceall}
\end{figure*}

The mean abundances and ionisation imbalances, $\imbalance$, are presented in
Table \ref{tablestarresults} and \fig{ionimbalanceall}.  The different stars are
discussed individually below.

The differences from \citet{Sitnova_2020} are small in LTE, but significant
enough to be noted. This could be attributed in part to the difference in the
microturbulence parameters in these stars. In this work, they are found to be
0.5\,${\rm km/s}$ 
smaller for HD84937, the same for HD140283, and 0.2\,${\rm km/s}$ larger for HD122563. Indeed, it is found that the discrepancy 
in LTE abundances
with respect to \citet{Sitnova_2020}
is largest for HD84937 and HD122563.
(Table \ref{tablestarresults}).

\subsubsection{The Sun}
\label{results_sun}

In the Sun, the 1D LTE abundances are found to be $4.87\pm0.01$ for \ion{Ti}{I}
and $4.92\pm0.01$ for \ion{Ti}{II}, where the uncertainties reflect the standard
error in the mean of the set of diagnostic lines. They are in excellent
agreement with the results from \citet{scott}, who obtained 4.85 and 4.91, respectively, in LTE using 1D \marcs{} models.  They are also consistent within
errors with the 1D LTE results from \citet{Sitnova_2020}, 4.88 and 4.95 respectively.

With the inclusion of 1D non-LTE effects, the abundances are found to be
$4.90\pm0.01$ and $4.92\pm0.01$ for \ion{Ti}{I} and \ion{Ti}{II} respectively.
This is again close to the corresponding abundances in
\citet{Sitnova_2020}, 4.91 and 4.94 respectively. 

The 1D LTE abundances correspond to an imbalance of $\imbalance=-0.06\pm0.05$ dex.
This is small, but significant, especially given that the solar parameters are
known precisely.  The 1D non-LTE abundances correspond to a much smaller
ionisation imbalance, $\imbalance=-0.02\pm0.05$ dex.  These imbalances are slightly smaller
in both magnitude and error than those of \citet{Sitnova_2020}.

\subsubsection{Arcturus}
\label{results_arcturus}

For Arcturus, the non-LTE effects for this mildly metal-poor star are small, as was also found for the Sun
(\sect{results_sun}). The 1D LTE titanium abundance
is found to be 4.82$\pm0.02$ from \ion{Ti}{I} lines,
and 4.93$\pm0.01$ from \ion{Ti}{II} lines.
In non-LTE, they change to 4.83$\pm0.03$ for \ion{Ti}{I} and 4.90$\pm0.00$ for \ion{Ti}{II}. These are higher than the \cite{arcturus} abundances of 4.66 for both ions, which can in part be explained by the higher microturbulence adopted in that work, and their line selection.

In 1D LTE, Arcturus has a marginally significant
titanium ionisation imbalance of $\imbalance =
-0.08\pm0.08$ dex. In non-LTE, the result is $\imbalance = -0.05\pm0.08$ dex.

\subsubsection{Very metal-poor giant: HD122563}

For HD122563, the 1D LTE titanium abundances are $2.09\pm0.03$ when inferred
from \ion{Ti}{I} lines, and $2.54\pm0.02$ from \ion{Ti}{II} lines.
These values are somewhat lower than the 1D LTE values
of \citet{Sitnova_2020}, who
find 2.19 for \ion{Ti}{I} and 2.58 for \ion{Ti}{II}.

The red giant suffers from the largest non-LTE effects:  for \ion{Ti}{I} they amount to a correction of $+0.32$ dex, such that the titanium abundance increases to $2.41\pm0.02$. For \ion{Ti}{II} the overall correction is just $+0.03$ dex, giving $2.57\pm0.02$ for \ion{Ti}{II}. The effects are much larger than those in Arcturus (\sect{results_arcturus}) due to the much lower metallicity of HD122563, which leads to greater over-ionisation.

Although the non-LTE correction on \ion{Ti}{I} found here is larger than that found by \citet{Sitnova_2020} ($+0.21$ dex), it turns out that the 1D non-LTE abundances from that work and this one are in good agreement; \citet{Sitnova_2020} find 2.40 for \ion{Ti}{I} and 2.60 for \ion{Ti}{II}.

The 1D LTE ionisation imbalance $\imbalance$ is found to also be the most severe of all the stars in this sample: $\imbalance = -0.45\pm0.08$ dex. The large non-LTE correction for \ion{Ti}{I}, and the correspondingly small correction for \ion{Ti}{II}, thus greatly improves this: $\imbalance = -0.16\pm0.08$ dex. Hence, it is found that, although 1D non-LTE improves the ionisation imbalance, it does not completely remove it. This motivates further study beyond 1D non-LTE for red giants. In other words, 3D non-LTE calculations.

\subsubsection{Very metal-poor dwarf and subgiant: HD84937 and HD140283}

In 1D LTE, the titanium abundance of
HD84937 is measured to be $3.20\pm0.01$ from \ion{Ti}{I} lines and 
$3.20\pm0.01$ from \ion{Ti}{II} lines.
In 1D non-LTE, the \ion{Ti}{I} result increases
by $+0.20$ dex to $3.40\pm0.01$,
while the \ion{Ti}{II} result is almost unchanged at 
$3.21\pm0.01$.
The correction for \ion{Ti}{I} is
larger than the $+0.12$ dex found in \citet{Sitnova_2020}.
Nevertheless, the 1D non-LTE
abundances found here are consistent with those in that work,
at $3.35$ and $3.18$ from
\ion{Ti}{I} and \ion{Ti}{II} lines, respectively.

The findings for the sub-giant
HD140283 are qualitatively similar,
with a large 1D non-LTE correction of $0.22$ dex for \ion{Ti}{I} lines
and a smaller one of $0.02$ dex for \ion{Ti}{II} lines.
The 1D non-LTE abundances found here, $2.91\pm0.01$
and $2.71\pm0.01$ respectively, are roughly consistent 
with those 
derived by \citet{Sitnova_2020}, 2.84 and 2.73, respectively.

Titanium ionisation balance is perfectly achieved
in 1D LTE for both 
HD84937 and HD140283: $\imbalance = 0.00\pm0.05$ dex.
Since the \ion{Ti}{I} lines suffer a considerable non-LTE effect
while the \ion{Ti}{II} lines are largely unaffected,
this means that a significant ionisation imbalance develops
in 1D non-LTE:
$\imbalance = 0.18\pm0.05$
and $0.20\pm0.05$ dex for the two stars, respectively.
This was also found in the independent study of
\citet{Sitnova_2020}: 1D non-LTE ionisation 
imbalances of
$\imbalance = 0.17$ and $0.11$ dex were found for these two stars,
whereas the ionisation balance was reduced to $0.05\,\mathrm{dex}$ or 
better in 1D LTE.

%------------------------------------------

\section{Conclusion}
\label{conclusions}

   Titanium abundances in late-type stars have been investigated using 1D model atmospheres and LTE and non-LTE radiative transfer. The present work makes use of an extended model atom that includes new quantum data for inelastic collisions with neutral hydrogen. Promisingly, 1D non-LTE models significantly reduce the ionisation imbalance, $\imbalance$, for the Sun (from $-0.06\pm0.05$ to $-0.02\pm0.05$ dex) and the very metal-poor giant HD122563 (from $-0.47\pm0.08$ to $-0.17\pm0.08$ dex), relative to 1D LTE.  At the same time, however, 1D non-LTE models worsen the ionisation imbalance for the very metal-poor stars HD84937 and HD140283 (by around $0.2$ dex). Titanium lines in the mildly metal-poor giant Arcturus were found to form very close to LTE, for which there remains a small ionisation imbalance ($-0.08\pm0.08$ dex). 
   
   These overall conclusions are broadly consistent with what was previously found by \citet{Sitnova_2020}, although the non-LTE effects are slightly larger for \ion{Ti}{I} and smaller for \ion{Ti}{II}.  This is reassuring, given that the current work uses a different radiative transfer code, \balder{}, and a more extended model atom that employs newer data for inelastic \ion{Ti}{I} collisions with neutral hydrogen.  
   
   It is quite possible that the residual ionisation imbalances for titanium are in part driven by 3D effects, which are significant for the Sun \citep{scott}. The effect can be expected to be even larger for metal-poor stars, for which the steep temperature gradients should enhance the non-LTE over-ionisation effects  \citep{anishFe,nordlander}.

   Further insight may be gained via a comparison to Fe, for which 1D and 3D non-LTE calculations have been performed. Several studies have found reasonable ionisation balances for HD84937 and HD140283 in 1D non-LTE using either the Drawin formula \citep{ironnlte} or asymptotic models such as those employed in this work \citep{anishFe}. However, for HD122563, \citet{anishFe} report an ionisation imbalance of around $\Delta_{\rm{Fe\,I-Fe\,II}} = 0.3$ dex in both 1D and 3D non-LTE, similar to the imbalance found in this work for titanium. This may point to as of yet unidentified shortcomings for metal-poor giants, but a larger sample of stars should be analysed before firm conclusions can be drawn. In any case, 3D non-LTE calculations for titanium will aid in the understanding of these problems.

\begin{acknowledgements}
JM and KL acknowledge funds from the European Research Council (ERC) under the European Union’s Horizon 2020 research and innovation programme (Grant agreement No. 852977). AMA, JG  and PSB acknowledge support from the Swedish Research Council through individual project grants with contract Nos. 2020-03940, 2020-05467 and 2020-03404. This work has made use of the VALD database, operated at Uppsala University, the Institute of Astronomy RAS in Moscow, and the University of Vienna. AKB gratefully acknowledges support from the Russian Science Foundation (the Russian Federation), Project No 22-23-01181.

\end{acknowledgements}
\bibliographystyle{aa}
\bibliography{references}

\begin{thebibliography}{67}
\expandafter\ifx\csname natexlab\endcsname\relax\def\natexlab#1{#1}\fi

\bibitem[{{Amarsi} {et~al.}(2018{\natexlab{a}}){Amarsi}, {Barklem, P. S.},
  {Asplund, M.}, {Collet, R.}, \& {Zatsarinny, O.}}]{anishOH}
{Amarsi}, A.~M., {Barklem, P. S.}, {Asplund, M.}, {Collet, R.}, \& {Zatsarinny,
  O.} 2018{\natexlab{a}}, A\&A, 616, A89

\bibitem[{{Amarsi} {et~al.}(2019){Amarsi}, {Barklem, P. S.}, {Collet, R.},
  {Grevesse, N.}, \& {Asplund, M.}}]{anishC}
{Amarsi}, A.~M., {Barklem, P. S.}, {Collet, R.}, {Grevesse, N.}, \& {Asplund,
  M.} 2019, A\&A, 624, A111

\bibitem[{{Amarsi} {et~al.}(2016){Amarsi}, {Lind}, {Asplund}, {Barklem}, \&
  {Collet}}]{anishFe}
{Amarsi}, A.~M., {Lind}, K., {Asplund}, M., {Barklem}, P.~S., \& {Collet}, R.
  2016, \mnras, 463, 1518

\bibitem[{{Amarsi} {et~al.}(2018{\natexlab{b}}){Amarsi}, {Nordlander, T.},
  {Barklem, P. S.}, {Asplund, M.}, {Collet, R.}, \& {Lind, K.}}]{balder}
{Amarsi}, A.~M., {Nordlander, T.}, {Barklem, P. S.}, {et~al.}
  2018{\natexlab{b}}, A\&A, 615, A139

\bibitem[{{Andrievsky} {et~al.}(2018){Andrievsky}, {Bonifacio}, {Caffau},
  {Korotin}, {Spite}, {Spite}, {Sbordone}, \& {Zhukova}}]{andrievskyCu}
{Andrievsky}, S., {Bonifacio}, P., {Caffau}, E., {et~al.} 2018, \mnras, 473,
  3377

\bibitem[{Barklem {et~al.}(2011)Barklem, Belyaev, Guitou, Feautrier, Gadéa, \&
  Spielfiedel}]{drawin}
Barklem, P., Belyaev, A., Guitou, M., {et~al.} 2011, Astronomy \& Astrophysics
  - ASTRON ASTROPHYS, 530

\bibitem[{Barklem(2016)}]{barklem2016b}
Barklem, P.~S. 2016, Phys. Rev. A, 93, 042705

\bibitem[{Barklem {et~al.}(2003)Barklem, Belyaev, \&
  Asplund}]{Barklem2003InelasticHA}
Barklem, P.~S., Belyaev, A.~K., \& Asplund, M. 2003, Astronomy and
  Astrophysics, 409

\bibitem[{{Barklem} {et~al.}(2010){Barklem}, {Belyaev}, {Dickinson}, \&
  {Gad{\'e}a}}]{barklemsodium}
{Barklem}, P.~S., {Belyaev}, A.~K., {Dickinson}, A.~S., \& {Gad{\'e}a}, F.~X.
  2010, \aap, 519, A20

\bibitem[{{Barklem} {et~al.}(2012){Barklem}, {Belyaev}, {Spielfiedel},
  {Guitou}, \& {Feautrier}}]{barklemmagnesium}
{Barklem}, P.~S., {Belyaev}, A.~K., {Spielfiedel}, A., {Guitou}, M., \&
  {Feautrier}, N. 2012, \aap, 541, A80

\bibitem[{Belyaev {et~al.}(2019)Belyaev, Vlasov, Mitrushchenkov, \&
  Feautrier}]{andreyCa}
Belyaev, A., Vlasov, D., Mitrushchenkov, A., \& Feautrier, N. 2019, Monthly
  Notices of the Royal Astronomical Society, 490, 3384

\bibitem[{Belyaev(2013)}]{bely2013}
Belyaev, A.~K. 2013, Phys. Rev. A, 88, 052704

\bibitem[{Belyaev \& Barklem(2003)}]{bely2003}
Belyaev, A.~K. \& Barklem, P.~S. 2003, Phys. Rev. A, 68, 062703

\bibitem[{Belyaev {et~al.}(2010)Belyaev, Barklem, Dickinson, \&
  Gad\'ea}]{bely2010}
Belyaev, A.~K., Barklem, P.~S., Dickinson, A.~S., \& Gad\'ea, F.~X. 2010, Phys.
  Rev. A, 81, 032706

\bibitem[{{Belyaev} {et~al.}(2012){Belyaev}, {Barklem}, {Spielfiedel},
  {Guitou}, {Feautrier}, {Rodionov}, \& {Vlasov}}]{AndreyMg}
{Belyaev}, A.~K., {Barklem}, P.~S., {Spielfiedel}, A., {et~al.} 2012, \pra, 85,
  032704

\bibitem[{Belyaev {et~al.}(1999)Belyaev, Grosser, Hahne, \& Menzel}]{bely1999}
Belyaev, A.~K., Grosser, J., Hahne, J., \& Menzel, T. 1999, Phys. Rev. A, 60,
  2151

\bibitem[{{Belyaev} \& {Voronov}(2018)}]{belyVoronov2018Simplifiedmodel}
{Belyaev}, A.~K. \& {Voronov}, Y.~V. 2018, \apj, 868, 86

\bibitem[{{Belyaev} {et~al.}(2017){Belyaev}, {Yakovleva}, \&
  {Kraemer}}]{belyetal2017SimplifiedModel}
{Belyaev}, A.~K., {Yakovleva}, S.~A., \& {Kraemer}, W.~P. 2017, European
  Physical Journal D, 71, 276

\bibitem[{{Bensby} {et~al.}(2014){Bensby}, {Feltzing, S.}, \& {Oey, M.
  S.}}]{disks}
{Bensby}, T., {Feltzing, S.}, \& {Oey, M. S.} 2014, A\&A, 562, A71

\bibitem[{Bergemann(2011)}]{berge}
Bergemann, M. 2011, Monthly Notices of The Royal Astronomical Society - MON
  NOTIC ROY ASTRON SOC, 413

\bibitem[{Bergemann \& Gehren(2008)}]{bergemann_Mn}
Bergemann, M. \& Gehren, T. 2008, Astronomy \& Astrophysics, 492, 823

\bibitem[{{Buder} {et~al.}(2021){Buder}, {Sharma}, {Kos}, {Amarsi},
  {Nordlander}, {Lind}, {Martell}, {Asplund}, {Bland-Hawthorn}, {Casey}, {de
  Silva}, {D'Orazi}, {Freeman}, {Hayden}, {Lewis}, {Lin}, {Schlesinger},
  {Simpson}, {Stello}, {Zucker}, {Zwitter}, {Beeson}, {Buck}, {Casagrande},
  {Clark}, {{\v{C}}otar}, {da Costa}, {de Grijs}, {Feuillet}, {Horner},
  {Kafle}, {Khanna}, {Kobayashi}, {Liu}, {Montet}, {Nandakumar}, {Nataf},
  {Ness}, {Spina}, {Tepper-Garc{\'\i}a}, {Ting}, {Traven},
  {Vogrin{\v{c}}i{\v{c}}}, {Wittenmyer}, {Wyse}, {{\v{Z}}erjal}, \& {GALAH
  Collaboration}}]{Buder21}
{Buder}, S., {Sharma}, S., {Kos}, J., {et~al.} 2021, \mnras, 506, 150

\bibitem[{Cox(2000)}]{allen}
Cox, A.~N. 2000, Allen's astrophysical quantities (New York : AIP Press :
  Springer)

\bibitem[{{Dalton} {et~al.}(2018){Dalton}, {Trager}, {Abrams}, {Bonifacio},
  {Aguerri}, {Vallenari}, {Middleton}, {Benn}, {Dee}, {Say{\`e}de}, {Lewis},
  {Pragt}, {Pic{\'o}}, {Walton}, {Rey}, {Allende}, {Lhom{\'e}}, {Terrett},
  {Brock}, {Gilbert}, {Ridings}, {Verheijen}, {Tosh}, {Steele}, {Stuik},
  {Kroes}, {Tromp}, {Kragt}, {Lesman}, {Mottram}, {Bates}, {Gribbin}, {Burgal},
  {Herreros}, {Delgado}, {Martin}, {Cano}, {Navarro}, {Irwin}, {Lewis},
  {Gonzales Solares}, {O'Mahony}, {Bianco}, {Zurita}, {ter Horst}, {Molinari},
  {Lodi}, {Guerra}, {Baruffolo}, {Carrasco}, {Farkas}, {Schallig}, {Hill},
  {Smith}, {Drew}, {Poggianti}, {Pieri}, {Jin}, {Dominquez Palmero},
  {Fari{\~n}a}, {Martin}, {Worley}, {Murphy}, {Hidalgo}, {Mignot}, {Bishop},
  {Guest}, {Elswijk}, {de Haan}, {Hanenburg}, {Salasnich}, {Mayya},
  {Izazaga-P{\'e}rez}, \& {Peralta de Arriba}}]{Dalton18}
{Dalton}, G., {Trager}, S., {Abrams}, D.~C., {et~al.} 2018, in Society of
  Photo-Optical Instrumentation Engineers (SPIE) Conference Series, Vol. 10702,
  107021B

\bibitem[{{de Jong} {et~al.}(2019){de Jong}, {Agertz}, {Berbel}, {Aird},
  {Alexander}, {Amarsi}, {Anders}, {Andrae}, {Ansarinejad}, {Ansorge},
  {Antilogus}, {Anwand-Heerwart}, {Arentsen}, {Arnadottir}, {Asplund}, {Auger},
  {Azais}, {Baade}, {Baker}, {Baker}, {Balbinot}, {Baldry}, {Banerji},
  {Barden}, {Barklem}, {Barth{\'e}l{\'e}my-Mazot}, {Battistini}, {Bauer},
  {Bell}, {Bellido-Tirado}, {Bellstedt}, {Belokurov}, {Bensby}, {Bergemann},
  {Bestenlehner}, {Bielby}, {Bilicki}, {Blake}, {Bland-Hawthorn}, {Boeche},
  {Boland}, {Boller}, {Bongard}, {Bongiorno}, {Bonifacio}, {Boudon}, {Brooks},
  {Brown}, {Brown}, {Br{\"u}ggen}, {Brynnel}, {Brzeski}, {Buchert},
  {Buschkamp}, {Caffau}, {Caillier}, {Carrick}, {Casagrande}, {Case}, {Casey},
  {Cesarini}, {Cescutti}, {Chapuis}, {Chiappini}, {Childress}, {Christlieb},
  {Church}, {Cioni}, {Cluver}, {Colless}, {Collett}, {Comparat}, {Cooper},
  {Couch}, {Courbin}, {Croom}, {Croton}, {Daguis{\'e}}, {Dalton}, {Davies},
  {Davis}, {de Laverny}, {Deason}, {Dionies}, {Disseau}, {Doel}, {D{\"o}scher},
  {Driver}, {Dwelly}, {Eckert}, {Edge}, {Edvardsson}, {Youssoufi}, {Elhaddad},
  {Enke}, {Erfanianfar}, {Farrell}, {Fechner}, {Feiz}, {Feltzing}, {Ferreras},
  {Feuerstein}, {Feuillet}, {Finoguenov}, {Ford}, {Fotopoulou}, {Fouesneau},
  {Frenk}, {Frey}, {Gaessler}, {Geier}, {Gentile Fusillo}, {Gerhard},
  {Giannantonio}, {Giannone}, {Gibson}, {Gillingham},
  {Gonz{\'a}lez-Fern{\'a}ndez}, {Gonzalez-Solares}, {Gottloeber}, {Gould},
  {Grebel}, {Gueguen}, {Guiglion}, {Haehnelt}, {Hahn}, {Hansen}, {Hartman},
  {Hauptner}, {Hawkins}, {Haynes}, {Haynes}, {Heiter}, {Helmi}, {Aguayo},
  {Hewett}, {Hinton}, {Hobbs}, {Hoenig}, {Hofman}, {Hook}, {Hopgood},
  {Hopkins}, {Hourihane}, {Howes}, {Howlett}, {Huet}, {Irwin}, {Iwert},
  {Jablonka}, {Jahn}, {Jahnke}, {Jarno}, {Jin}, {Jofre}, {Johl}, {Jones},
  {J{\"o}nsson}, {Jordan}, {Karovicova}, {Khalatyan}, {Kelz}, {Kennicutt},
  {King}, {Kitaura}, {Klar}, {Klauser}, {Kneib}, {Koch}, {Koposov},
  {Kordopatis}, {Korn}, {Kosmalski}, {Kotak}, {Kovalev}, {Kreckel}, {Kripak},
  {Krumpe}, {Kuijken}, {Kunder}, {Kushniruk}, {Lam}, {Lamer}, {Laurent},
  {Lawrence}, {Lehmitz}, {Lemasle}, {Lewis}, {Li}, {Lidman}, {Lind}, {Liske},
  {Lizon}, {Loveday}, {Ludwig}, {McDermid}, {Maguire}, {Mainieri}, {Mali},
  {Mandel}, {Mandel}, {Mannering}, {Martell}, {Martinez Delgado}, {Matijevic},
  {McGregor}, {McMahon}, {McMillan}, {Mena}, {Merloni}, {Meyer}, {Michel},
  {Micheva}, {Migniau}, {Minchev}, {Monari}, {Muller}, {Murphy},
  {Muthukrishna}, {Nandra}, {Navarro}, {Ness}, {Nichani}, {Nichol}, {Nicklas},
  {Niederhofer}, {Norberg}, {Obreschkow}, {Oliver}, {Owers}, {Pai},
  {Pankratow}, {Parkinson}, {Paschke}, {Paterson}, {Pecontal}, {Parry},
  {Phillips}, {Pillepich}, {Pinard}, {Pirard}, {Piskunov}, {Plank},
  {Pl{\"u}schke}, {Pons}, {Popesso}, {Power}, {Pragt}, {Pramskiy}, {Pryer},
  {Quattri}, {Queiroz}, {Quirrenbach}, {Rahurkar}, {Raichoor}, {Ramstedt},
  {Rau}, {Recio-Blanco}, {Reiss}, {Renaud}, {Revaz}, {Rhode}, {Richard},
  {Richter}, {Rix}, {Robotham}, {Roelfsema}, {Romaniello}, {Rosario},
  {Rothmaier}, {Roukema}, {Ruchti}, {Rupprecht}, {Rybizki}, {Ryde}, {Saar},
  {Sadler}, {Sahl{\'e}n}, {Salvato}, {Sassolas}, {Saunders}, {Saviauk},
  {Sbordone}, {Schmidt}, {Schnurr}, {Scholz}, {Schwope}, {Seifert}, {Shanks},
  {Sheinis}, {Sivov}, {Sk{\'u}lad{\'o}ttir}, {Smartt}, {Smedley}, {Smith},
  {Smith}, {Sorce}, {Spitler}, {Starkenburg}, {Steinmetz}, {Stilz}, {Storm},
  {Sullivan}, {Sutherland}, {Swann}, {Tamone}, {Taylor}, {Teillon}, {Tempel},
  {ter Horst}, {Thi}, {Tolstoy}, {Trager}, {Traven}, {Tremblay}, {Tresse},
  {Valentini}, {van de Weygaert}, {van den Ancker}, {Veljanoski}, {Venkatesan},
  {Wagner}, {Wagner}, {Walcher}, {Waller}, {Walton}, {Wang}, {Winkler},
  {Wisotzki}, {Worley}, {Worseck}, {Xiang}, {Xu}, {Yong}, {Zhao}, {Zheng},
  {Zscheyge}, \& {Zucker}}]{deJong19}
{de Jong}, R.~S., {Agertz}, O., {Berbel}, A.~A., {et~al.} 2019, The Messenger,
  175, 3

\bibitem[{Drawin(1968)}]{drawin1968formelmassigen}
Drawin, H.-W. 1968, Zeitschrift f{\"u}r Physik A Hadrons and nuclei, 211, 404

\bibitem[{Drawin(1969)}]{drawin1969}
Drawin, H.~W. 1969, Z. Phys., 225: 483-93(1969).

\bibitem[{{Gaia Collaboration} {et~al.}(2016){Gaia Collaboration}, {Prusti},
  {de Bruijne}, {Brown}, {Vallenari}, {Babusiaux}, {Bailer-Jones}, {Bastian},
  {Biermann}, {Evans}, {Eyer}, {Jansen}, {Jordi}, {Klioner}, {Lammers},
  {Lindegren}, {Luri}, {Mignard}, {Milligan}, {Panem}, {Poinsignon},
  {Pourbaix}, {Randich}, {Sarri}, {Sartoretti}, {Siddiqui}, {Soubiran},
  {Valette}, {van Leeuwen}, {Walton}, {Aerts}, {Arenou}, {Cropper}, {Drimmel},
  {H{\o}g}, {Katz}, {Lattanzi}, {O'Mullane}, {Grebel}, {Holland}, {Huc},
  {Passot}, {Bramante}, {Cacciari}, {Casta{\~n}eda}, {Chaoul}, {Cheek}, {De
  Angeli}, {Fabricius}, {Guerra}, {Hern{\'a}ndez}, {Jean-Antoine-Piccolo},
  {Masana}, {Messineo}, {Mowlavi}, {Nienartowicz}, {Ord{\'o}{\~n}ez-Blanco},
  {Panuzzo}, {Portell}, {Richards}, {Riello}, {Seabroke}, {Tanga},
  {Th{\'e}venin}, {Torra}, {Els}, {Gracia-Abril}, {Comoretto},
  {Garcia-Reinaldos}, {Lock}, {Mercier}, {Altmann}, {Andrae}, {Astraatmadja},
  {Bellas-Velidis}, {Benson}, {Berthier}, {Blomme}, {Busso}, {Carry},
  {Cellino}, {Clementini}, {Cowell}, {Creevey}, {Cuypers}, {Davidson}, {De
  Ridder}, {de Torres}, {Delchambre}, {Dell'Oro}, {Ducourant}, {Fr{\'e}mat},
  {Garc{\'\i}a-Torres}, {Gosset}, {Halbwachs}, {Hambly}, {Harrison}, {Hauser},
  {Hestroffer}, {Hodgkin}, {Huckle}, {Hutton}, {Jasniewicz}, {Jordan},
  {Kontizas}, {Korn}, {Lanzafame}, {Manteiga}, {Moitinho}, {Muinonen},
  {Osinde}, {Pancino}, {Pauwels}, {Petit}, {Recio-Blanco}, {Robin}, {Sarro},
  {Siopis}, {Smith}, {Smith}, {Sozzetti}, {Thuillot}, {van Reeven}, {Viala},
  {Abbas}, {Abreu Aramburu}, {Accart}, {Aguado}, {Allan}, {Allasia},
  {Altavilla}, {{\'A}lvarez}, {Alves}, {Anderson}, {Andrei}, {Anglada Varela},
  {Antiche}, {Antoja}, {Ant{\'o}n}, {Arcay}, {Atzei}, {Ayache}, {Bach},
  {Baker}, {Balaguer-N{\'u}{\~n}ez}, {Barache}, {Barata}, {Barbier}, {Barblan},
  {Baroni}, {Barrado y Navascu{\'e}s}, {Barros}, {Barstow}, {Becciani},
  {Bellazzini}, {Bellei}, {Bello Garc{\'\i}a}, {Belokurov}, {Bendjoya},
  {Berihuete}, {Bianchi}, {Bienaym{\'e}}, {Billebaud}, {Blagorodnova},
  {Blanco-Cuaresma}, {Boch}, {Bombrun}, {Borrachero}, {Bouquillon}, {Bourda},
  {Bouy}, {Bragaglia}, {Breddels}, {Brouillet}, {Br{\"u}semeister},
  {Bucciarelli}, {Budnik}, {Burgess}, {Burgon}, {Burlacu}, {Busonero}, {Buzzi},
  {Caffau}, {Cambras}, {Campbell}, {Cancelliere}, {Cantat-Gaudin}, {Carlucci},
  {Carrasco}, {Castellani}, {Charlot}, {Charnas}, {Charvet}, {Chassat},
  {Chiavassa}, {Clotet}, {Cocozza}, {Collins}, {Collins}, {Costigan}, {Crifo},
  {Cross}, {Crosta}, {Crowley}, {Dafonte}, {Damerdji}, {Dapergolas}, {David},
  {David}, {De Cat}, {de Felice}, {de Laverny}, {De Luise}, {De March}, {de
  Martino}, {de Souza}, {Debosscher}, {del Pozo}, {Delbo}, {Delgado},
  {Delgado}, {di Marco}, {Di Matteo}, {Diakite}, {Distefano}, {Dolding}, {Dos
  Anjos}, {Drazinos}, {Dur{\'a}n}, {Dzigan}, {Ecale}, {Edvardsson}, {Enke},
  {Erdmann}, {Escolar}, {Espina}, {Evans}, {Eynard Bontemps}, {Fabre},
  {Fabrizio}, {Faigler}, {Falc{\~a}o}, {Farr{\`a}s Casas}, {Faye}, {Federici},
  {Fedorets}, {Fern{\'a}ndez-Hern{\'a}ndez}, {Fernique}, {Fienga}, {Figueras},
  {Filippi}, {Findeisen}, {Fonti}, {Fouesneau}, {Fraile}, {Fraser}, {Fuchs},
  {Furnell}, {Gai}, {Galleti}, {Galluccio}, {Garabato}, {Garc{\'\i}a-Sedano},
  {Gar{\'e}}, {Garofalo}, {Garralda}, {Gavras}, {Gerssen}, {Geyer}, {Gilmore},
  {Girona}, {Giuffrida}, {Gomes}, {Gonz{\'a}lez-Marcos},
  {Gonz{\'a}lez-N{\'u}{\~n}ez}, {Gonz{\'a}lez-Vidal}, {Granvik}, {Guerrier},
  {Guillout}, {Guiraud}, {G{\'u}rpide}, {Guti{\'e}rrez-S{\'a}nchez}, {Guy},
  {Haigron}, {Hatzidimitriou}, {Haywood}, {Heiter}, {Helmi}, {Hobbs},
  {Hofmann}, {Holl}, {Holland}, {Hunt}, {Hypki}, {Icardi}, {Irwin}, {Jevardat
  de Fombelle}, {Jofr{\'e}}, {Jonker}, {Jorissen}, {Julbe}, {Karampelas},
  {Kochoska}, {Kohley}, {Kolenberg}, {Kontizas}, {Koposov}, {Kordopatis},
  {Koubsky}, {Kowalczyk}, {Krone-Martins}, {Kudryashova}, {Kull}, {Bachchan},
  {Lacoste-Seris}, {Lanza}, {Lavigne}, {Le Poncin-Lafitte}, {Lebreton},
  {Lebzelter}, {Leccia}, {Leclerc}, {Lecoeur-Taibi}, {Lemaitre}, {Lenhardt},
  {Leroux}, {Liao}, {Licata}, {Lindstr{\o}m}, {Lister}, {Livanou}, {Lobel},
  {L{\"o}ffler}, {L{\'o}pez}, {Lopez-Lozano}, {Lorenz}, {Loureiro},
  {MacDonald}, {Magalh{\~a}es Fernandes}, {Managau}, {Mann}, {Mantelet},
  {Marchal}, {Marchant}, {Marconi}, {Marie}, {Marinoni}, {Marrese},
  {Marschalk{\'o}}, {Marshall}, {Mart{\'\i}n-Fleitas}, {Martino}, {Mary},
  {Matijevi{\v{c}}}, {Mazeh}, {McMillan}, {Messina}, {Mestre}, {Michalik},
  {Millar}, {Miranda}, {Molina}, {Molinaro}, {Molinaro}, {Moln{\'a}r},
  {Moniez}, {Montegriffo}, {Monteiro}, {Mor}, {Mora}, {Morbidelli}, {Morel},
  {Morgenthaler}, {Morley}, {Morris}, {Mulone}, {Muraveva}, {Musella},
  {Narbonne}, {Nelemans}, {Nicastro}, {Noval}, {Ord{\'e}novic},
  {Ordieres-Mer{\'e}}, {Osborne}, {Pagani}, {Pagano}, {Pailler}, {Palacin},
  {Palaversa}, {Parsons}, {Paulsen}, {Pecoraro}, {Pedrosa}, {Pentik{\"a}inen},
  {Pereira}, {Pichon}, {Piersimoni}, {Pineau}, {Plachy}, {Plum}, {Poujoulet},
  {Pr{\v{s}}a}, {Pulone}, {Ragaini}, {Rago}, {Rambaux}, {Ramos-Lerate},
  {Ranalli}, {Rauw}, {Read}, {Regibo}, {Renk}, {Reyl{\'e}}, {Ribeiro},
  {Rimoldini}, {Ripepi}, {Riva}, {Rixon}, {Roelens}, {Romero-G{\'o}mez},
  {Rowell}, {Royer}, {Rudolph}, {Ruiz-Dern}, {Sadowski}, {Sagrist{\`a}
  Sell{\'e}s}, {Sahlmann}, {Salgado}, {Salguero}, {Sarasso}, {Savietto},
  {Schnorhk}, {Schultheis}, {Sciacca}, {Segol}, {Segovia}, {Segransan},
  {Serpell}, {Shih}, {Smareglia}, {Smart}, {Smith}, {Solano}, {Solitro},
  {Sordo}, {Soria Nieto}, {Souchay}, {Spagna}, {Spoto}, {Stampa}, {Steele},
  {Steidelm{\"u}ller}, {Stephenson}, {Stoev}, {Suess}, {S{\"u}veges}, {Surdej},
  {Szabados}, {Szegedi-Elek}, {Tapiador}, {Taris}, {Tauran}, {Taylor},
  {Teixeira}, {Terrett}, {Tingley}, {Trager}, {Turon}, {Ulla}, {Utrilla},
  {Valentini}, {van Elteren}, {Van Hemelryck}, {van Leeuwen}, {Varadi},
  {Vecchiato}, {Veljanoski}, {Via}, {Vicente}, {Vogt}, {Voss}, {Votruba},
  {Voutsinas}, {Walmsley}, {Weiler}, {Weingrill}, {Werner}, {Wevers},
  {Whitehead}, {Wyrzykowski}, {Yoldas}, {{\v{Z}}erjal}, {Zucker}, {Zurbach},
  {Zwitter}, {Alecu}, {Allen}, {Allende Prieto}, {Amorim},
  {Anglada-Escud{\'e}}, {Arsenijevic}, {Azaz}, {Balm}, {Beck}, {Bernstein},
  {Bigot}, {Bijaoui}, {Blasco}, {Bonfigli}, {Bono}, {Boudreault}, {Bressan},
  {Brown}, {Brunet}, {Bunclark}, {Buonanno}, {Butkevich}, {Carret}, {Carrion},
  {Chemin}, {Ch{\'e}reau}, {Corcione}, {Darmigny}, {de Boer}, {de Teodoro}, {de
  Zeeuw}, {Delle Luche}, {Domingues}, {Dubath}, {Fodor}, {Fr{\'e}zouls},
  {Fries}, {Fustes}, {Fyfe}, {Gallardo}, {Gallegos}, {Gardiol}, {Gebran},
  {Gomboc}, {G{\'o}mez}, {Grux}, {Gueguen}, {Heyrovsky}, {Hoar}, {Iannicola},
  {Isasi Parache}, {Janotto}, {Joliet}, {Jonckheere}, {Keil}, {Kim},
  {Klagyivik}, {Klar}, {Knude}, {Kochukhov}, {Kolka}, {Kos}, {Kutka}, {Lainey},
  {LeBouquin}, {Liu}, {Loreggia}, {Makarov}, {Marseille}, {Martayan},
  {Martinez-Rubi}, {Massart}, {Meynadier}, {Mignot}, {Munari}, {Nguyen},
  {Nordlander}, {Ocvirk}, {O'Flaherty}, {Olias Sanz}, {Ortiz}, {Osorio},
  {Oszkiewicz}, {Ouzounis}, {Palmer}, {Park}, {Pasquato}, {Peltzer}, {Peralta},
  {P{\'e}turaud}, {Pieniluoma}, {Pigozzi}, {Poels}, {Prat}, {Prod'homme},
  {Raison}, {Rebordao}, {Risquez}, {Rocca-Volmerange}, {Rosen}, {Ruiz-Fuertes},
  {Russo}, {Sembay}, {Serraller Vizcaino}, {Short}, {Siebert}, {Silva},
  {Sinachopoulos}, {Slezak}, {Soffel}, {Sosnowska}, {Strai{\v{z}}ys}, {ter
  Linden}, {Terrell}, {Theil}, {Tiede}, {Troisi}, {Tsalmantza}, {Tur},
  {Vaccari}, {Vachier}, {Valles}, {Van Hamme}, {Veltz}, {Virtanen}, {Wallut},
  {Wichmann}, {Wilkinson}, {Ziaeepour}, \& {Zschocke}}]{Gaia2016}
{Gaia Collaboration}, {Prusti}, T., {de Bruijne}, J.~H.~J., {et~al.} 2016,
  \aap, 595, A1

\bibitem[{Grumer \& Barklem(2020)}]{grumer}
Grumer, J. \& Barklem, P.~S. 2020, Astronomy \& Astrophysics, 637, A28

\bibitem[{Guitou {et~al.}(2015)Guitou, Spielfiedel, Rodionov, Yakovleva,
  Belyaev, Merle, Thévenin, \& Feautrier}]{chemicalMg}
Guitou, M., Spielfiedel, A., Rodionov, D., {et~al.} 2015, Chemical Physics,
  462, 94

\bibitem[{{Gustafsson} {et~al.}(2008){Gustafsson}, {Edvardsson, B.}, {Eriksson,
  K.}, {J\o{}rgensen, U. G.}, {Nordlund, \AA{}.}, \& {Plez,
  B.}}]{MARCSdatabase}
{Gustafsson}, B., {Edvardsson, B.}, {Eriksson, K.}, {et~al.} 2008, A\&A, 486,
  951

\bibitem[{{Heiter} {et~al.}(2015){Heiter}, {Jofr{\'e}}, {Gustafsson}, {Korn},
  {Soubiran}, \& {Th{\'e}venin}}]{startableb}
{Heiter}, U., {Jofr{\'e}}, P., {Gustafsson}, B., {et~al.} 2015, \aap, 582, A49

\bibitem[{{Helmi}(2020)}]{Helmi20}
{Helmi}, A. 2020, \araa, 58, 205

\bibitem[{{Karovicova} {et~al.}(2020){Karovicova}, {White}, {Nordlander},
  {Casagrande}, {Ireland}, {Huber}, \& {Jofr{\'e}}}]{startablec}
{Karovicova}, I., {White}, T.~R., {Nordlander}, T., {et~al.} 2020, \aap, 640,
  A25

\bibitem[{Kaulakys(1991)}]{kaulakys}
Kaulakys, B. 1991, 24, L127

\bibitem[{{Kurucz}(2016)}]{K16}
{Kurucz}, R.~L. 2016, Robert L. Kurucz on-line database of observed and
  predicted atomic transitions

\bibitem[{{Lambert}(1993)}]{lambert_hydrogen}
{Lambert}, D.~L. 1993, Physica Scripta Volume T, 47, 186

\bibitem[{{Lawler} {et~al.}(2013){Lawler}, {Guzman}, {Wood}, {Sneden}, \&
  {Cowan}}]{HD84937-2}
{Lawler}, J.~E., {Guzman}, A., {Wood}, M.~P., {Sneden}, C., \& {Cowan}, J.~J.
  2013, \apjs, 205, 11

\bibitem[{{Leenaarts} \& {Carlsson}(2009)}]{leenaart}
{Leenaarts}, J. \& {Carlsson}, M. 2009, in Astronomical Society of the Pacific
  Conference Series, Vol. 415, The Second Hinode Science Meeting: Beyond
  Discovery-Toward Understanding, ed. B.~{Lites}, M.~{Cheung}, T.~{Magara},
  J.~{Mariska}, \& K.~{Reeves}, 87

\bibitem[{{Lind} {et~al.}(2017){Lind}, {Amarsi}, {Asplund}, {Barklem},
  {Bautista}, {Bergemann}, {Collet}, {Kiselman}, {Leenaarts}, \&
  {Pereira}}]{lind17}
{Lind}, K., {Amarsi}, A.~M., {Asplund}, M., {et~al.} 2017, \mnras, 468, 4311

\bibitem[{{Lind} {et~al.}(2009){Lind}, {Asplund, M.}, \& {Barklem, P.
  S.}}]{lindbarklem2009}
{Lind}, K., {Asplund, M.}, \& {Barklem, P. S.} 2009, A\&A, 503, 541

\bibitem[{Lind {et~al.}(2022)Lind, Nordlander, Wehrhahn, Montelius, Osorio,
  Barklem, Afsar, Sneden, \& Kobayashi}]{KarinParameters}
Lind, K., Nordlander, T., Wehrhahn, A., {et~al.} 2022, Non-LTE abundance
  corrections for late-type stars from 2000$\AA$ to 3$\mu$m: I. Na, Mg, and Al

\bibitem[{{Majewski} {et~al.}(2017){Majewski}, {Schiavon}, {Frinchaboy},
  {Allende Prieto}, {Barkhouser}, {Bizyaev}, {Blank}, {Brunner}, {Burton},
  {Carrera}, {Chojnowski}, {Cunha}, {Epstein}, {Fitzgerald}, {Garc{\'\i}a
  P{\'e}rez}, {Hearty}, {Henderson}, {Holtzman}, {Johnson}, {Lam}, {Lawler},
  {Maseman}, {M{\'e}sz{\'a}ros}, {Nelson}, {Nguyen}, {Nidever}, {Pinsonneault},
  {Shetrone}, {Smee}, {Smith}, {Stolberg}, {Skrutskie}, {Walker}, {Wilson},
  {Zasowski}, {Anders}, {Basu}, {Beland}, {Blanton}, {Bovy}, {Brownstein},
  {Carlberg}, {Chaplin}, {Chiappini}, {Eisenstein}, {Elsworth}, {Feuillet},
  {Fleming}, {Galbraith-Frew}, {Garc{\'\i}a}, {Garc{\'\i}a-Hern{\'a}ndez},
  {Gillespie}, {Girardi}, {Gunn}, {Hasselquist}, {Hayden}, {Hekker}, {Ivans},
  {Kinemuchi}, {Klaene}, {Mahadevan}, {Mathur}, {Mosser}, {Muna}, {Munn},
  {Nichol}, {O'Connell}, {Parejko}, {Robin}, {Rocha-Pinto}, {Schultheis},
  {Serenelli}, {Shane}, {Silva Aguirre}, {Sobeck}, {Thompson}, {Troup},
  {Weinberg}, \& {Zamora}}]{Majewski17}
{Majewski}, S.~R., {Schiavon}, R.~P., {Frinchaboy}, P.~M., {et~al.} 2017, \aj,
  154, 94

\bibitem[{{Masseron}(2006)}]{MasseronPHD}
{Masseron}, T. 2006, PhD thesis, Obs. de Paris

\bibitem[{{Mihalas}(1978)}]{hydrogenic}
{Mihalas}, D. 1978, {Stellar atmospheres}

\bibitem[{Nahar(2015)}]{nahar2015}
Nahar, S. 2015, New Astronomy, 38

\bibitem[{{Nahar}(2020)}]{nahardatabse}
{Nahar}, S. 2020, Atoms, 8, 68

\bibitem[{{Nissen} {et~al.}(2020){Nissen}, {Christensen-Dalsgaard},
  {Mosumgaard}, {Silva Aguirre}, {Spitoni}, \& {Verma}}]{stellarage}
{Nissen}, P.~E., {Christensen-Dalsgaard}, J., {Mosumgaard}, J.~R., {et~al.}
  2020, \aap, 640, A81

\bibitem[{{Nissen} \& {Gustafsson}(2018)}]{Nissen18}
{Nissen}, P.~E. \& {Gustafsson}, B. 2018, \aapr, 26, 6

\bibitem[{{Nordlander} {et~al.}(2017){Nordlander}, {Amarsi}, {Lind}, {Asplund},
  {Barklem}, {Casey}, {Collet}, \& {Leenaarts}}]{nordlander}
{Nordlander}, T., {Amarsi}, A.~M., {Lind}, K., {et~al.} 2017, \aap, 597, A6

\bibitem[{Olson {et~al.}(1971)Olson, Smith, \& Bauer}]{olson1971estimation}
Olson, R., Smith, F., \& Bauer, E. 1971, Applied Optics, 10, 1848

\bibitem[{{Osorio} {et~al.}(2015){Osorio}, {Barklem, P. S.}, {Lind, K.},
  {Belyaev, A. K.}, {Spielfiedel, A.}, {Guitou, M.}, \& {Feautrier,
  N.}}]{osorio2015}
{Osorio}, Y., {Barklem, P. S.}, {Lind, K.}, {et~al.} 2015, A\&A, 579, A53

\bibitem[{{Przybilla} {et~al.}(2011){Przybilla}, {Nieva}, \&
  {Butler}}]{detailcode}
{Przybilla}, N., {Nieva}, M.-F., \& {Butler}, K. 2011, in Journal of Physics
  Conference Series, Vol. 328, Journal of Physics Conference Series, 012015

\bibitem[{Prša {et~al.}(2016)Prša, Harmanec, Torres, Mamajek, Asplund,
  Capitaine, Christensen-Dalsgaard, Depagne, Haberreiter, Hekker, \&
  et~al.}]{startablea}
Prša, A., Harmanec, P., Torres, G., {et~al.} 2016, The Astronomical Journal,
  152, 41

\bibitem[{Ramírez \& Allende~Prieto(2011)}]{arcturus}
Ramírez, I. \& Allende~Prieto, C. 2011, The Astrophysical Journal, 743, 135

\bibitem[{{Reggiani} {et~al.}(2019){Reggiani}, {Amarsi, Anish M.}, {Lind,
  Karin}, {Barklem, Paul S.}, {Zatsarinny, Oleg}, {Bartschat, Klaus}, {Fursa,
  Dmitry V.}, {Bray, Igor}, {Spina, Lorenzo}, \& {Mel\'endez,
  Jorge}}]{reggiani2019}
{Reggiani}, H., {Amarsi, Anish M.}, {Lind, Karin}, {et~al.} 2019, A\&A, 627,
  A177

\bibitem[{{Ryabchikova} {et~al.}(2015){Ryabchikova}, {Piskunov}, {Kurucz},
  {Stempels}, {Heiter}, {Pakhomov}, \& {Barklem}}]{vald}
{Ryabchikova}, T., {Piskunov}, N., {Kurucz}, R.~L., {et~al.} 2015, \physscr,
  90, 054005

\bibitem[{{Scott} {et~al.}(2015){Scott}, {Asplund, Martin}, {Grevesse,
  Nicolas}, {Bergemann, Maria}, \& {Jacques Sauval, A.}}]{scott}
{Scott}, P., {Asplund, Martin}, {Grevesse, Nicolas}, {Bergemann, Maria}, \&
  {Jacques Sauval, A.} 2015, A\&A, 573, A26

\bibitem[{{Shi} {et~al.}(2018){Shi}, {Yan}, {Zhou}, \& {Zhao}}]{shiCu}
{Shi}, J.~R., {Yan}, H.~L., {Zhou}, Z.~M., \& {Zhao}, G. 2018, \apj, 862, 71

\bibitem[{Sitnova {et~al.}(2016)Sitnova, Mashonkina, \&
  Ryabchikova}]{sitnova2016}
Sitnova, T.~M., Mashonkina, L.~I., \& Ryabchikova, T.~A. 2016, Monthly Notices
  of the Royal Astronomical Society, 461, 1000

\bibitem[{{Sitnova} {et~al.}(2020){Sitnova}, {Yakovleva}, {Belyaev}, \&
  {Mashonkina}}]{Sitnova_2020}
{Sitnova}, T.~M., {Yakovleva}, S.~A., {Belyaev}, A.~K., \& {Mashonkina}, L.~I.
  2020, Astronomy Letters, 46, 120

\bibitem[{Sitnova {et~al.}(2022)Sitnova, Yakovleva, Belyaev, \&
  Mashonkina}]{sitnova_Zn}
Sitnova, T.~M., Yakovleva, S.~A., Belyaev, A.~K., \& Mashonkina, L.~I. 2022,
  Monthly Notices of the Royal Astronomical Society, 515, 1510

\bibitem[{Steenbock \& Holweger(1984)}]{steenbock1984statistical}
Steenbock, W. \& Holweger, H. 1984, Astronomy and Astrophysics, 130, 319

\bibitem[{{van Regemorter}(1962)}]{regemorter}
{van Regemorter}, H. 1962, \apj, 136, 906

\bibitem[{Wood {et~al.}(2013)Wood, Lawler, Sneden, \& Cowan}]{HD84937}
Wood, M.~P., Lawler, J.~E., Sneden, C., \& Cowan, J.~J. 2013, The Astrophysical
  Journal Supplement Series, 208, 27

\bibitem[{{Zhao} {et~al.}(2016){Zhao}, {Mashonkina}, {Yan}, {Alexeeva},
  {Kobayashi}, {Pakhomov}, {Shi}, {Sitnova}, {Tan}, {Zhang}, {Zhang}, {Zhou},
  {Bolte}, {Chen}, {Li}, {Liu}, \& {Zhai}}]{ironnlte}
{Zhao}, G., {Mashonkina}, L., {Yan}, H.~L., {et~al.} 2016, \apj, 833, 225

\bibitem[{{Zhao} {et~al.}(2012){Zhao}, {Zhao}, {Chu}, {Jing}, \&
  {Deng}}]{Zhao12}
{Zhao}, G., {Zhao}, Y.-H., {Chu}, Y.-Q., {Jing}, Y.-P., \& {Deng}, L.-C. 2012,
  Research in Astronomy and Astrophysics, 12, 723

\end{thebibliography}

\onecolumn
\begin{appendix}

\section{Additional table}
\label{appx}

This section contains the linelists used for both \ion{Ti}{I} and \ion{Ti}{II} in each star individually, as well as the necessary information on each transition in the model atom.

\begin{longtable}{r r r r r r r}
\caption{Lines considered in the analysis, including saturated lines that were removed during final abundance calculation. Errors for equivalent widths were not available for the lines of the Sun from \citet{scott}.}
\label{appxtable}
\\
\hline
\noalign{\smallskip}

Wavelength & $E_{exc}$ & log($gf$) & $W_{\lambda}$ & Error & \multicolumn{2}{c}{Abundances}\\
\r{A}  & $\rm{eV}$   & & m\r{A} & & LTE & NLTE  \\ 
\noalign{\smallskip}
\hline
\hline
\noalign{\smallskip}

\endhead

\noalign{\smallskip}
\multicolumn{7}{c}{ Sun }\\
\noalign{\smallskip}
\noalign{\smallskip}
\multicolumn{7}{c}{ \ion{Ti}{I} }\\
\noalign{\smallskip}
4281.367 & 0.813 & -1.260 & 24.000 & --- & 4.84 & 4.87 \\
4465.805 & 1.739 & -0.130 & 35.600 & --- & 4.84 & 4.88 \\
4758.118 & 2.249 & 0.510 & 41.800 & --- & 4.78 & 4.83 \\
4759.270 & 2.256 & 0.590 & 46.000 & --- & 4.79 & 4.84 \\
5022.868 & 0.826 & -0.330 & 69.900 & --- & 4.78 & 4.78 \\
5113.440 & 1.443 & -0.700 & 24.500 & --- & 4.82 & 4.85 \\
5145.460 & 1.460 & -0.540 & 33.100 & --- & 4.86 & 4.89 \\
5147.478 & 0.000 & -1.940 & 34.600 & --- & 4.86 & 4.90 \\
5152.184 & 0.021 & -1.950 & 33.200 & --- & 4.86 & 4.90 \\
5219.702 & 0.021 & -2.220 & 22.400 & --- & 4.89 & 4.93 \\
5252.100 & 0.048 & -2.360 & 16.400 & --- & 4.88 & 4.92 \\
5295.776 & 1.067 & -1.590 & 10.600 & --- & 4.89 & 4.93 \\
5490.148 & 1.460 & -0.840 & 20.300 & --- & 4.84 & 4.87 \\
6092.792 & 1.887 & -1.380 & 3.600 & --- & 4.92 & 4.96 \\
6258.102 & 1.443 & -0.390 & 50.500 & --- & 4.92 & 4.94 \\
6303.757 & 1.443 & -1.580 & 6.800 & --- & 4.97 & 5.00 \\
6312.236 & 1.460 & -1.550 & 6.800 & --- & 4.95 & 4.98 \\
7357.727 & 1.443 & -1.020 & 19.900 & --- & 4.87 & 4.89 \\
8675.372 & 1.067 & -1.500 & 18.500 & --- & 4.86 & 4.89 \\
8682.983 & 1.053 & -1.790 & 10.700 & --- & 4.86 & 4.89 \\
8692.329 & 1.046 & -2.130 & 5.200 & --- & 4.86 & 4.89 \\
8734.710 & 1.053 & -2.240 & 4.100 & --- & 4.87 & 4.90 \\
\noalign{\smallskip}
\multicolumn{7}{c}{ \ion{Ti}{II} }\\
\noalign{\smallskip}
4409.518 & 1.231 & -2.530 & 38.100 & --- & 4.90 & 4.89 \\
4444.554 & 1.116 & -2.200 & 59.900 & --- & 4.90 & 4.90 \\
4493.522 & 1.080 & -2.780 & 31.800 & --- & 4.86 & 4.86 \\
4583.409 & 1.165 & -2.840 & 30.200 & --- & 4.96 & 4.96 \\
4609.265 & 1.180 & -3.320 & 11.600 & --- & 4.91 & 4.91 \\
4657.201 & 1.243 & -2.290 & 51.800 & --- & 4.93 & 4.92 \\
4708.663 & 1.237 & -2.350 & 50.600 & --- & 4.95 & 4.95 \\
4719.511 & 1.243 & -3.320 & 12.200 & --- & 4.99 & 4.99 \\
4764.525 & 1.237 & -2.690 & 33.500 & --- & 4.93 & 4.93 \\
4798.531 & 1.080 & -2.660 & 42.900 & --- & 4.94 & 4.94 \\
4865.610 & 1.116 & -2.700 & 35.000 & --- & 4.85 & 4.85 \\
5336.786 & 1.582 & -1.600 & 72.000 & --- & 4.91 & 4.90 \\
5381.022 & 1.566 & -1.970 & 56.600 & --- & 4.94 & 4.93 \\
5418.768 & 1.582 & -2.130 & 48.100 & --- & 4.94 & 4.94 \\
\noalign{\smallskip}
\hline
\noalign{\smallskip}
\multicolumn{7}{c}{ 84937 }\\
\noalign{\smallskip}
\noalign{\smallskip}
\multicolumn{7}{c}{ \ion{Ti}{I} }\\
\noalign{\smallskip}
2646.634 & 0.048 & 0.060 & 16.723 & 0.881 & 3.22 & 3.45 \\
2956.132 & 0.048 & 0.120 & 15.147 & 0.708 & 3.05 & 3.30 \\
3186.451 & 0.000 & 0.010 & 18.269 & 0.249 & 3.19 & 3.44 \\
3191.992 & 0.021 & 0.160 & 23.646 & 0.293 & 3.22 & 3.47 \\
3309.496 & 1.053 & -0.190 & 11.933 & 0.190 & 3.14 & 3.18 \\
3354.633 & 0.021 & 0.110 & 20.764 & 0.327 & 3.18 & 3.42 \\
3370.434 & 0.000 & -0.400 & 7.455 & 0.183 & 3.10 & 3.36 \\
3371.452 & 0.048 & 0.230 & 24.541 & 0.311 & 3.18 & 3.42 \\
3385.941 & 0.048 & -0.180 & 10.698 & 0.197 & 3.11 & 3.37 \\
3635.462 & 0.000 & 0.100 & 20.818 & 0.243 & 3.06 & 3.30 \\
3729.807 & 0.000 & -0.280 & 12.278 & 0.169 & 3.11 & 3.33 \\
3741.059 & 0.021 & -0.150 & 15.099 & 0.198 & 3.10 & 3.33 \\
3904.783 & 0.900 & 0.150 & 6.148 & 0.145 & 3.13 & 3.32 \\
3924.526 & 0.021 & -0.870 & 4.474 & 0.137 & 3.21 & 3.43 \\
3947.768 & 0.021 & -0.890 & 4.528 & 0.133 & 3.23 & 3.45 \\
3958.205 & 0.048 & -0.110 & 21.996 & 0.241 & 3.28 & 3.49 \\
3989.758 & 0.021 & -0.130 & 20.710 & 0.235 & 3.24 & 3.45 \\
3998.636 & 0.048 & 0.020 & 22.842 & 0.249 & 3.17 & 3.40 \\
4008.927 & 0.021 & -1.000 & 4.184 & 0.133 & 3.30 & 3.51 \\
4024.571 & 0.048 & -0.920 & 4.603 & 0.338 & 3.29 & 3.50 \\
4025.077 & 2.153 & -1.040 & 19.407 & 0.203 & 3.21 & 3.21 \\
4287.403 & 0.836 & -0.370 & 2.993 & 0.116 & 3.23 & 3.36 \\
4305.907 & 0.848 & 0.490 & 20.131 & 0.231 & 3.32 & 3.43 \\
4427.098 & 1.502 & 0.230 & 2.837 & 0.180 & 3.20 & 3.37 \\
4449.142 & 1.887 & 0.470 & 2.994 & 0.129 & 3.33 & 3.47 \\
4518.022 & 0.826 & -0.250 & 3.747 & 0.127 & 3.19 & 3.32 \\
4533.239 & 0.848 & 0.540 & 17.596 & 0.200 & 3.18 & 3.31 \\
4534.776 & 0.836 & 0.350 & 12.076 & 0.158 & 3.16 & 3.29 \\
4535.569 & 0.826 & 0.140 & 9.021 & 0.143 & 3.21 & 3.34 \\
4548.764 & 0.826 & -0.280 & 4.395 & 0.144 & 3.29 & 3.41 \\
4555.483 & 0.848 & -0.400 & 2.536 & 0.115 & 3.18 & 3.31 \\
4617.269 & 1.749 & 0.440 & 2.982 & 0.111 & 3.22 & 3.38 \\
4981.731 & 0.848 & 0.570 & 20.401 & 0.190 & 3.21 & 3.34 \\
4991.066 & 0.836 & 0.450 & 16.413 & 0.361 & 3.20 & 3.32 \\
4999.503 & 0.826 & 0.320 & 13.348 & 0.168 & 3.21 & 3.33 \\
5022.868 & 0.826 & -0.330 & 2.989 & 0.115 & 3.14 & 3.27 \\
5036.464 & 1.443 & 0.140 & 4.104 & 0.117 & 3.37 & 3.53 \\
5173.743 & 0.000 & -1.060 & 3.627 & 0.111 & 3.20 & 3.45 \\
5192.969 & 0.021 & -0.950 & 4.682 & 0.113 & 3.23 & 3.47 \\
\noalign{\smallskip}
\multicolumn{7}{c}{ \ion{Ti}{II} }\\
\noalign{\smallskip}
2474.194 & 0.049 & -2.420 & 22.823 & 1.818 & 3.44 & 3.45 \\
2517.431 & 0.135 & -1.500 & 53.129 & 4.940 & 3.58 & 3.58 \\
2571.032 & 0.607 & -0.900 & 62.659 & 7.945 & 3.74 & 3.70 \\
2581.711 & 1.084 & -1.580 & 22.415 & 1.166 & 3.49 & 3.50 \\
2717.297 & 1.130 & -1.490 & 26.340 & 5.347 & 3.52 & 3.52 \\
2725.773 & 1.116 & -1.550 & 20.050 & 1.534 & 3.40 & 3.41 \\
2761.287 & 1.080 & -1.350 & 23.567 & 0.923 & 3.27 & 3.29 \\
2784.638 & 0.607 & -1.990 & 11.026 & 0.900 & 3.03 & 3.04 \\
2820.361 & 0.574 & -1.910 & 21.286 & 0.834 & 3.29 & 3.30 \\
2832.176 & 0.574 & -0.850 & 65.543 & 2.287 & 3.65 & 3.62 \\
2841.935 & 0.607 & -0.590 & 58.541 & 3.441 & 3.19 & 3.19 \\
2862.319 & 1.237 & -0.530 & 49.528 & 4.582 & 3.35 & 3.37 \\
2884.102 & 1.130 & -0.230 & 60.614 & 1.318 & 3.35 & 3.32 \\
3017.183 & 1.584 & -0.300 & 48.744 & 0.907 & 3.36 & 3.36 \\
3029.728 & 1.572 & -0.350 & 38.428 & 0.968 & 3.08 & 3.11 \\
3046.684 & 1.165 & -0.810 & 40.677 & 1.121 & 3.23 & 3.25 \\
3056.738 & 1.161 & -0.790 & 42.627 & 1.063 & 3.27 & 3.28 \\
3058.088 & 1.180 & -0.420 & 54.899 & 1.022 & 3.32 & 3.31 \\
3071.239 & 1.180 & -0.750 & 52.036 & 0.632 & 3.49 & 3.49 \\
3089.400 & 1.893 & 0.080 & 42.717 & 0.454 & 3.06 & 3.12 \\
3103.803 & 1.892 & 0.180 & 51.629 & 0.712 & 3.24 & 3.30 \\
3105.080 & 1.224 & -0.430 & 52.530 & 0.787 & 3.27 & 3.27 \\
3106.231 & 1.243 & -0.070 & 56.670 & 0.543 & 3.07 & 3.08 \\
3110.080 & 1.582 & -1.210 & 12.605 & 1.576 & 3.16 & 3.19 \\
3117.666 & 1.231 & -0.490 & 41.388 & 1.879 & 2.98 & 3.01 \\
3122.070 & 1.237 & -1.570 & 11.697 & 0.540 & 3.16 & 3.18 \\
3144.719 & 0.113 & -2.360 & 24.452 & 1.897 & 3.35 & 3.35 \\
3154.192 & 0.113 & -1.160 & 69.833 & 0.771 & 3.56 & 3.49 \\
3184.117 & 0.012 & -2.520 & 18.383 & 0.716 & 3.23 & 3.23 \\
3197.519 & 0.028 & -1.940 & 41.150 & 0.474 & 3.30 & 3.28 \\
3203.431 & 0.000 & -1.820 & 52.780 & 1.610 & 3.48 & 3.44 \\
3213.121 & 0.012 & -2.280 & 26.246 & 0.873 & 3.21 & 3.21 \\
3214.767 & 0.049 & -1.370 & 67.276 & 0.971 & 3.60 & 3.51 \\
3224.237 & 1.584 & 0.050 & 53.705 & 0.554 & 3.12 & 3.12 \\
3226.769 & 0.028 & -1.790 & 52.173 & 0.718 & 3.46 & 3.42 \\
3236.119 & 1.080 & -0.410 & 53.626 & 5.112 & 3.07 & 3.10 \\
3249.366 & 1.080 & -1.350 & 27.618 & 0.442 & 3.30 & 3.32 \\
3263.683 & 1.165 & -1.140 & 26.920 & 0.394 & 3.15 & 3.18 \\
3272.077 & 1.224 & -0.250 & 51.744 & 1.237 & 3.02 & 3.06 \\
3275.290 & 1.080 & -1.480 & 16.020 & 0.224 & 3.08 & 3.11 \\
3278.288 & 1.231 & -0.260 & 58.420 & 0.502 & 3.26 & 3.28 \\
3279.988 & 1.116 & -1.190 & 28.995 & 0.699 & 3.21 & 3.23 \\
3282.327 & 1.224 & -0.340 & 52.539 & 0.561 & 3.14 & 3.16 \\
3302.095 & 0.151 & -2.340 & 20.723 & 0.304 & 3.24 & 3.23 \\
3307.721 & 0.122 & -2.660 & 12.661 & 0.204 & 3.25 & 3.24 \\
3308.803 & 0.135 & -1.240 & 68.140 & 0.518 & 3.54 & 3.49 \\
3315.322 & 1.224 & -0.640 & 44.200 & 0.578 & 3.17 & 3.20 \\
3318.023 & 0.122 & -1.070 & 73.533 & 0.530 & 3.55 & 3.49 \\
3319.081 & 0.135 & -3.000 & 7.573 & 0.178 & 3.33 & 3.33 \\
3337.847 & 1.237 & -1.250 & 19.774 & 0.279 & 3.11 & 3.14 \\
3343.762 & 0.151 & -1.180 & 69.415 & 0.495 & 3.54 & 3.48 \\
3352.069 & 1.221 & -1.280 & 22.323 & 0.298 & 3.20 & 3.23 \\
3369.203 & 1.231 & -1.420 & 15.098 & 0.212 & 3.12 & 3.14 \\
3374.346 & 1.237 & -1.060 & 28.908 & 3.204 & 3.18 & 3.20 \\
3388.751 & 1.237 & -1.020 & 27.381 & 0.364 & 3.10 & 3.13 \\
3407.202 & 0.049 & -1.970 & 39.785 & 0.969 & 3.28 & 3.27 \\
3409.808 & 0.028 & -1.910 & 43.641 & 0.506 & 3.30 & 3.29 \\
3416.957 & 1.237 & -1.540 & 11.353 & 0.230 & 3.09 & 3.12 \\
3452.465 & 2.048 & -0.560 & 15.914 & 0.262 & 3.03 & 3.10 \\
3456.384 & 2.061 & -0.110 & 30.175 & 0.368 & 3.01 & 3.11 \\
3461.496 & 0.135 & -0.850 & 82.020 & 7.164 & 3.60 & 3.49 \\
3477.180 & 0.122 & -0.950 & 80.637 & 1.814 & 3.63 & 3.53 \\
3489.736 & 0.135 & -2.000 & 42.603 & 0.469 & 3.43 & 3.42 \\
3491.049 & 0.113 & -1.100 & 74.271 & 0.491 & 3.56 & 3.47 \\
3500.333 & 0.122 & -2.040 & 32.866 & 0.354 & 3.22 & 3.22 \\
3504.891 & 1.892 & 0.380 & 56.694 & 1.011 & 3.12 & 3.11 \\
3520.252 & 2.048 & -0.180 & 26.988 & 0.310 & 2.98 & 3.05 \\
3533.854 & 2.061 & -1.310 & 2.634 & 0.151 & 2.91 & 2.96 \\
3535.407 & 2.061 & 0.010 & 32.222 & 0.321 & 2.94 & 3.02 \\
3561.576 & 0.574 & -2.040 & 17.707 & 0.465 & 3.21 & 3.21 \\
3573.732 & 0.574 & -1.530 & 35.625 & 1.841 & 3.20 & 3.19 \\
3596.047 & 0.607 & -1.070 & 53.092 & 0.638 & 3.26 & 3.21 \\
3741.638 & 1.582 & -0.070 & 59.641 & 0.677 & 3.11 & 3.14 \\
3757.685 & 1.566 & -0.440 & 45.258 & 0.503 & 3.08 & 3.12 \\
3759.291 & 0.607 & 0.280 & 118.779 & 0.718 & 3.35 & 3.25 \\
3761.321 & 0.574 & 0.180 & 112.735 & 0.673 & 3.32 & 3.22 \\
3761.872 & 2.590 & -0.420 & 9.989 & 0.157 & 2.97 & 3.01 \\
3786.323 & 0.607 & -2.600 & 8.811 & 0.556 & 3.31 & 3.31 \\
3813.388 & 0.607 & -1.890 & 27.090 & 0.289 & 3.22 & 3.22 \\
3900.539 & 1.130 & -0.290 & 80.301 & 0.528 & 3.49 & 3.38 \\
3987.606 & 0.607 & -2.730 & 5.331 & 0.142 & 3.18 & 3.18 \\
4012.384 & 0.574 & -1.780 & 38.560 & 0.369 & 3.32 & 3.31 \\
4028.338 & 1.892 & -0.920 & 17.673 & 0.216 & 3.14 & 3.17 \\
4053.821 & 1.893 & -1.070 & 13.115 & 0.171 & 3.12 & 3.15 \\
4161.529 & 1.084 & -2.090 & 8.613 & 0.148 & 3.19 & 3.19 \\
4163.644 & 2.590 & -0.130 & 23.253 & 0.242 & 3.13 & 3.17 \\
4171.904 & 2.598 & -0.300 & 18.053 & 0.208 & 3.15 & 3.20 \\
4290.215 & 1.165 & -0.870 & 53.161 & 0.593 & 3.27 & 3.24 \\
4300.043 & 1.180 & -0.460 & 72.296 & 0.782 & 3.39 & 3.31 \\
4301.923 & 1.161 & -1.210 & 38.264 & 0.418 & 3.25 & 3.25 \\
4312.860 & 1.180 & -1.120 & 42.831 & 0.503 & 3.28 & 3.27 \\
4316.794 & 2.048 & -1.620 & 3.773 & 0.122 & 3.19 & 3.23 \\
4320.950 & 1.165 & -1.880 & 13.552 & 0.406 & 3.27 & 3.27 \\
4391.026 & 1.231 & -2.300 & 6.717 & 0.141 & 3.39 & 3.39 \\
4395.839 & 1.243 & -1.930 & 9.203 & 0.146 & 3.18 & 3.19 \\
4409.518 & 1.231 & -2.530 & 2.082 & 0.116 & 3.09 & 3.10 \\
4418.331 & 1.237 & -1.990 & 8.415 & 0.142 & 3.20 & 3.20 \\
4443.801 & 1.080 & -0.710 & 69.175 & 0.448 & 3.43 & 3.33 \\
4444.554 & 1.116 & -2.200 & 6.948 & 0.131 & 3.20 & 3.20 \\
4468.493 & 1.130 & -0.630 & 71.117 & 0.462 & 3.45 & 3.34 \\
4488.324 & 3.123 & -0.500 & 4.765 & 0.133 & 3.14 & 3.18 \\
4493.522 & 1.080 & -2.780 & 1.934 & 0.111 & 3.16 & 3.16 \\
4501.270 & 1.116 & -0.770 & 65.839 & 0.474 & 3.42 & 3.34 \\
4518.332 & 1.080 & -2.560 & 4.121 & 0.132 & 3.28 & 3.28 \\
4545.133 & 1.130 & -2.450 & 3.986 & 0.133 & 3.20 & 3.20 \\
4571.971 & 1.572 & -0.310 & 63.803 & 0.423 & 3.33 & 3.25 \\
4583.409 & 1.165 & -2.840 & 1.789 & 0.103 & 3.26 & 3.26 \\
4657.201 & 1.243 & -2.290 & 4.909 & 0.147 & 3.23 & 3.23 \\
4708.663 & 1.237 & -2.350 & 4.197 & 0.114 & 3.21 & 3.21 \\
4762.778 & 1.084 & -2.890 & 1.443 & 0.119 & 3.13 & 3.13 \\
4763.883 & 1.221 & -2.400 & 4.295 & 0.120 & 3.25 & 3.26 \\
4764.525 & 1.237 & -2.690 & 2.066 & 0.106 & 3.23 & 3.23 \\
4798.531 & 1.080 & -2.660 & 2.612 & 0.198 & 3.16 & 3.16 \\
4911.194 & 3.123 & -0.640 & 3.448 & 0.110 & 3.11 & 3.19 \\
5013.686 & 1.582 & -2.140 & 4.012 & 0.136 & 3.28 & 3.28 \\
5129.156 & 1.892 & -1.340 & 9.976 & 0.154 & 3.19 & 3.18 \\
5185.902 & 1.893 & -1.410 & 8.391 & 0.136 & 3.17 & 3.17 \\
5211.530 & 2.590 & -1.410 & 1.205 & 0.097 & 2.92 & 2.99 \\
5336.786 & 1.582 & -1.600 & 10.452 & 0.138 & 3.18 & 3.17 \\
5381.022 & 1.566 & -1.970 & 5.299 & 0.141 & 3.20 & 3.20 \\
5418.768 & 1.582 & -2.130 & 3.676 & 0.153 & 3.21 & 3.20 \\
\noalign{\smallskip}
\hline
\noalign{\smallskip}
\multicolumn{7}{c}{ 140283 }\\
\noalign{\smallskip}
\noalign{\smallskip}
\multicolumn{7}{c}{ \ion{Ti}{I} }\\
\noalign{\smallskip}
3981.762 & 0.000 & -0.270 & 25.044 & 0.268 & 2.92 & 3.14 \\
2646.634 & 0.048 & 0.060 & 21.293 & 0.479 & 2.81 & 3.06 \\
3191.992 & 0.021 & 0.160 & 25.781 & 0.429 & 2.69 & 2.97 \\
3199.914 & 0.048 & 0.310 & 28.752 & 0.341 & 2.65 & 2.94 \\
3309.496 & 1.053 & -0.190 & 12.151 & 0.208 & 2.72 & 2.77 \\
3354.633 & 0.021 & 0.110 & 22.144 & 0.310 & 2.62 & 2.90 \\
3370.434 & 0.000 & -0.400 & 9.251 & 0.198 & 2.62 & 2.89 \\
3371.452 & 0.048 & 0.230 & 25.988 & 0.317 & 2.62 & 2.90 \\
3385.941 & 0.048 & -0.180 & 11.835 & 0.210 & 2.57 & 2.85 \\
3635.462 & 0.000 & 0.100 & 24.615 & 0.323 & 2.60 & 2.86 \\
3729.807 & 0.000 & -0.280 & 14.543 & 0.190 & 2.65 & 2.90 \\
3741.059 & 0.021 & -0.150 & 18.645 & 0.246 & 2.67 & 2.92 \\
3904.783 & 0.900 & 0.150 & 6.980 & 0.147 & 2.72 & 2.95 \\
3924.526 & 0.021 & -0.870 & 4.706 & 0.152 & 2.69 & 2.94 \\
3947.768 & 0.021 & -0.890 & 5.038 & 0.150 & 2.74 & 2.98 \\
3958.205 & 0.048 & -0.110 & 22.101 & 0.221 & 2.74 & 3.00 \\
3989.758 & 0.021 & -0.130 & 22.249 & 0.317 & 2.73 & 2.98 \\
3998.636 & 0.048 & 0.020 & 25.204 & 0.273 & 2.68 & 2.94 \\
4008.927 & 0.021 & -1.000 & 4.274 & 0.133 & 2.76 & 3.01 \\
4024.571 & 0.048 & -0.920 & 3.899 & 0.172 & 2.67 & 2.92 \\
4025.077 & 2.153 & -1.040 & 18.395 & 0.199 & 2.75 & 2.75 \\
4287.403 & 0.836 & -0.370 & 1.681 & 0.107 & 2.50 & 2.65 \\
4449.142 & 1.887 & 0.470 & 2.543 & 0.107 & 2.86 & 3.02 \\
4518.022 & 0.826 & -0.250 & 2.945 & 0.116 & 2.60 & 2.75 \\
4533.239 & 0.848 & 0.540 & 16.912 & 0.177 & 2.68 & 2.82 \\
4534.776 & 0.836 & 0.350 & 11.320 & 0.158 & 2.64 & 2.79 \\
4535.569 & 0.826 & 0.140 & 7.941 & 0.142 & 2.67 & 2.81 \\
4548.764 & 0.826 & -0.280 & 3.542 & 0.119 & 2.71 & 2.85 \\
4555.483 & 0.848 & -0.400 & 2.533 & 0.108 & 2.70 & 2.85 \\
4617.269 & 1.749 & 0.440 & 2.450 & 0.101 & 2.73 & 2.90 \\
4981.731 & 0.848 & 0.570 & 18.916 & 0.195 & 2.69 & 2.83 \\
4991.066 & 0.836 & 0.450 & 15.790 & 0.229 & 2.69 & 2.84 \\
4999.503 & 0.826 & 0.320 & 12.664 & 0.166 & 2.69 & 2.84 \\
5022.868 & 0.826 & -0.330 & 3.346 & 0.096 & 2.71 & 2.86 \\
5173.743 & 0.000 & -1.060 & 3.636 & 0.102 & 2.65 & 2.93 \\
5192.969 & 0.021 & -0.950 & 5.308 & 0.105 & 2.73 & 3.00 \\
\noalign{\smallskip}
\multicolumn{7}{c}{ \ion{Ti}{II} }\\
\noalign{\smallskip}
2474.194 & 0.049 & -2.420 & 16.588 & 0.479 & 2.77 & 2.78 \\
2571.032 & 0.607 & -0.900 & 49.930 & 1.408 & 2.82 & 2.82 \\
2725.773 & 1.116 & -1.550 & 16.545 & 0.478 & 2.88 & 2.89 \\
2761.287 & 1.080 & -1.350 & 17.913 & 0.362 & 2.69 & 2.71 \\
2784.638 & 0.607 & -1.990 & 11.018 & 0.339 & 2.59 & 2.60 \\
2832.176 & 0.574 & -0.850 & 73.872 & 0.669 & 3.40 & 3.38 \\
2841.935 & 0.607 & -0.590 & 57.579 & 0.633 & 2.63 & 2.66 \\
2862.319 & 1.237 & -0.530 & 39.353 & 0.691 & 2.60 & 2.65 \\
2884.102 & 1.130 & -0.230 & 53.627 & 0.743 & 2.62 & 2.64 \\
2888.929 & 0.574 & -1.360 & 35.823 & 1.749 & 2.67 & 2.70 \\
3029.728 & 1.572 & -0.350 & 35.125 & 0.496 & 2.59 & 2.63 \\
3046.684 & 1.165 & -0.810 & 33.755 & 0.442 & 2.61 & 2.65 \\
3056.738 & 1.161 & -0.790 & 39.140 & 0.490 & 2.72 & 2.75 \\
3058.088 & 1.180 & -0.420 & 46.472 & 0.431 & 2.58 & 2.62 \\
3089.400 & 1.893 & 0.080 & 36.812 & 0.592 & 2.51 & 2.57 \\
3103.803 & 1.892 & 0.180 & 39.238 & 0.476 & 2.46 & 2.54 \\
3105.080 & 1.224 & -0.430 & 45.552 & 0.652 & 2.59 & 2.63 \\
3106.231 & 1.243 & -0.070 & 56.597 & 0.684 & 2.59 & 2.63 \\
3110.080 & 1.582 & -1.210 & 9.683 & 0.737 & 2.64 & 2.68 \\
3154.192 & 0.113 & -1.160 & 74.321 & 0.566 & 3.12 & 3.08 \\
3184.117 & 0.012 & -2.520 & 20.083 & 0.383 & 2.78 & 2.77 \\
3195.715 & 1.084 & -1.390 & 17.194 & 0.538 & 2.62 & 2.67 \\
3197.519 & 0.028 & -1.940 & 43.688 & 1.099 & 2.83 & 2.82 \\
3203.431 & 0.000 & -1.820 & 52.338 & 1.250 & 2.93 & 2.90 \\
3213.121 & 0.012 & -2.280 & 29.684 & 0.979 & 2.78 & 2.78 \\
3224.237 & 1.584 & 0.050 & 49.048 & 0.612 & 2.53 & 2.57 \\
3226.769 & 0.028 & -1.790 & 53.110 & 1.526 & 2.94 & 2.91 \\
3263.683 & 1.165 & -1.140 & 23.765 & 0.297 & 2.64 & 2.69 \\
3272.077 & 1.224 & -0.250 & 52.508 & 1.251 & 2.56 & 2.62 \\
3275.290 & 1.080 & -1.480 & 16.374 & 0.279 & 2.67 & 2.71 \\
3276.992 & 0.122 & -2.440 & 16.969 & 0.235 & 2.69 & 2.69 \\
3278.288 & 1.231 & -0.260 & 50.905 & 0.561 & 2.53 & 2.59 \\
3279.988 & 1.116 & -1.190 & 22.024 & 0.275 & 2.59 & 2.63 \\
3282.327 & 1.224 & -0.340 & 47.776 & 0.533 & 2.52 & 2.57 \\
3302.095 & 0.151 & -2.340 & 19.755 & 0.248 & 2.71 & 2.71 \\
3307.721 & 0.122 & -2.660 & 14.970 & 0.215 & 2.84 & 2.84 \\
3308.803 & 0.135 & -1.240 & 66.696 & 0.739 & 2.91 & 2.89 \\
3315.322 & 1.224 & -0.640 & 39.956 & 0.449 & 2.60 & 2.67 \\
3318.023 & 0.122 & -1.070 & 72.373 & 0.513 & 2.91 & 2.88 \\
3319.081 & 0.135 & -3.000 & 7.919 & 0.193 & 2.86 & 2.86 \\
3337.847 & 1.237 & -1.250 & 17.151 & 0.226 & 2.61 & 2.65 \\
3343.762 & 0.151 & -1.180 & 68.002 & 0.475 & 2.89 & 2.87 \\
3352.069 & 1.221 & -1.280 & 21.122 & 0.274 & 2.75 & 2.79 \\
3369.203 & 1.231 & -1.420 & 13.929 & 0.219 & 2.66 & 2.69 \\
3374.346 & 1.237 & -1.060 & 24.712 & 3.943 & 2.64 & 2.68 \\
3407.202 & 0.049 & -1.970 & 44.831 & 0.623 & 2.86 & 2.86 \\
3409.808 & 0.028 & -1.910 & 45.101 & 0.538 & 2.79 & 2.78 \\
3416.957 & 1.237 & -1.540 & 9.988 & 0.204 & 2.61 & 2.65 \\
3452.465 & 2.048 & -0.560 & 11.867 & 0.266 & 2.52 & 2.60 \\
3456.384 & 2.061 & -0.110 & 24.325 & 0.288 & 2.49 & 2.59 \\
3461.496 & 0.135 & -0.850 & 80.096 & 5.435 & 2.90 & 2.85 \\
3489.736 & 0.135 & -2.000 & 46.395 & 0.573 & 3.00 & 3.00 \\
3491.049 & 0.113 & -1.100 & 71.993 & 0.686 & 2.86 & 2.82 \\
3500.333 & 0.122 & -2.040 & 35.313 & 0.400 & 2.75 & 2.75 \\
3504.891 & 1.892 & 0.380 & 49.854 & 0.624 & 2.47 & 2.52 \\
3520.252 & 2.048 & -0.180 & 21.796 & 0.483 & 2.47 & 2.55 \\
3533.854 & 2.061 & -1.310 & 3.172 & 0.148 & 2.64 & 2.69 \\
3535.407 & 2.061 & 0.010 & 27.127 & 0.313 & 2.43 & 2.53 \\
3561.576 & 0.574 & -2.040 & 15.988 & 0.270 & 2.68 & 2.68 \\
3573.732 & 0.574 & -1.530 & 35.594 & 2.193 & 2.69 & 2.70 \\
3596.047 & 0.607 & -1.070 & 52.056 & 0.556 & 2.68 & 2.68 \\
3741.638 & 1.582 & -0.070 & 53.184 & 0.610 & 2.56 & 2.63 \\
3757.685 & 1.566 & -0.440 & 37.523 & 0.387 & 2.54 & 2.61 \\
3759.291 & 0.607 & 0.280 & 111.636 & 0.687 & 2.75 & 2.66 \\
3761.321 & 0.574 & 0.180 & 108.670 & 0.656 & 2.76 & 2.67 \\
3761.872 & 2.590 & -0.420 & 8.304 & 0.153 & 2.61 & 2.65 \\
3786.323 & 0.607 & -2.600 & 8.394 & 0.529 & 2.86 & 2.87 \\
3813.388 & 0.607 & -1.890 & 25.384 & 0.279 & 2.75 & 2.75 \\
3900.539 & 1.130 & -0.290 & 69.866 & 0.725 & 2.76 & 2.70 \\
3913.461 & 1.116 & -0.360 & 65.483 & 1.049 & 2.68 & 2.64 \\
3981.991 & 0.574 & -2.540 & 8.381 & 0.143 & 2.75 & 2.75 \\
3987.606 & 0.607 & -2.730 & 5.223 & 0.142 & 2.75 & 2.75 \\
4012.384 & 0.574 & -1.780 & 34.284 & 0.336 & 2.79 & 2.79 \\
4028.338 & 1.892 & -0.920 & 12.998 & 0.195 & 2.65 & 2.70 \\
4053.821 & 1.893 & -1.070 & 9.145 & 0.147 & 2.62 & 2.67 \\
4161.529 & 1.084 & -2.090 & 7.348 & 0.147 & 2.73 & 2.74 \\
4163.644 & 2.590 & -0.130 & 14.121 & 0.184 & 2.58 & 2.65 \\
4171.904 & 2.598 & -0.300 & 10.990 & 0.155 & 2.63 & 2.69 \\
4290.215 & 1.165 & -0.870 & 46.801 & 0.436 & 2.71 & 2.72 \\
4300.043 & 1.180 & -0.460 & 67.900 & 0.754 & 2.84 & 2.81 \\
4301.923 & 1.161 & -1.210 & 38.424 & 0.525 & 2.87 & 2.87 \\
4316.794 & 2.048 & -1.620 & 2.012 & 0.104 & 2.61 & 2.67 \\
4320.950 & 1.165 & -1.880 & 11.160 & 0.171 & 2.79 & 2.80 \\
4330.698 & 1.180 & -2.090 & 6.467 & 0.124 & 2.75 & 2.76 \\
4418.331 & 1.237 & -1.990 & 7.433 & 0.138 & 2.76 & 2.78 \\
4443.801 & 1.080 & -0.710 & 61.977 & 0.401 & 2.82 & 2.76 \\
4444.554 & 1.116 & -2.200 & 5.172 & 0.115 & 2.68 & 2.68 \\
4450.482 & 1.084 & -1.520 & 24.059 & 0.262 & 2.76 & 2.75 \\
4468.493 & 1.130 & -0.630 & 64.215 & 0.414 & 2.84 & 2.78 \\
4488.324 & 3.123 & -0.500 & 1.828 & 0.100 & 2.49 & 2.53 \\
4493.522 & 1.080 & -2.780 & 1.295 & 0.096 & 2.60 & 2.61 \\
4501.270 & 1.116 & -0.770 & 58.311 & 0.390 & 2.81 & 2.77 \\
4518.332 & 1.080 & -2.560 & 3.297 & 0.114 & 2.79 & 2.80 \\
4545.133 & 1.130 & -2.450 & 3.575 & 0.107 & 2.77 & 2.78 \\
4571.971 & 1.572 & -0.310 & 53.657 & 0.361 & 2.69 & 2.67 \\
4583.409 & 1.165 & -2.840 & 1.635 & 0.099 & 2.84 & 2.84 \\
4657.201 & 1.243 & -2.290 & 3.385 & 0.103 & 2.69 & 2.69 \\
4708.663 & 1.237 & -2.350 & 3.855 & 0.113 & 2.80 & 2.80 \\
4763.883 & 1.221 & -2.400 & 3.939 & 0.104 & 2.84 & 2.85 \\
4764.525 & 1.237 & -2.690 & 1.504 & 0.102 & 2.71 & 2.72 \\
4798.531 & 1.080 & -2.660 & 2.616 & 0.167 & 2.77 & 2.78 \\
5013.686 & 1.582 & -2.140 & 2.117 & 0.103 & 2.64 & 2.67 \\
5129.156 & 1.892 & -1.340 & 6.935 & 0.136 & 2.69 & 2.70 \\
5185.902 & 1.893 & -1.410 & 4.998 & 0.104 & 2.60 & 2.62 \\
5211.530 & 2.590 & -1.410 & 1.178 & 0.085 & 2.65 & 2.72 \\
5336.786 & 1.582 & -1.600 & 7.759 & 0.118 & 2.68 & 2.69 \\
5381.022 & 1.566 & -1.970 & 3.967 & 0.104 & 2.72 & 2.73 \\
5418.768 & 1.582 & -2.130 & 3.381 & 0.105 & 2.82 & 2.82 \\
\noalign{\smallskip}
\hline
\noalign{\smallskip}
\multicolumn{7}{c}{ Arcturus }\\
\noalign{\smallskip}
\noalign{\smallskip}
\multicolumn{7}{c}{ \ion{Ti}{I} }\\
\noalign{\smallskip}
4759.270 & 2.256 & 0.590 & 112.543 & 1.788 & 4.77 & 4.74 \\
4778.255 & 2.236 & -0.350 & 84.851 & 1.277 & 5.02 & 4.98 \\
4913.613 & 1.873 & 0.220 & 137.628 & 2.769 & 5.14 & 5.09 \\
4915.229 & 1.887 & -0.910 & 72.443 & 0.911 & 4.80 & 4.74 \\
4926.148 & 0.818 & -2.090 & 93.792 & 0.497 & 4.87 & 4.83 \\
4997.097 & 0.000 & -2.070 & 167.350 & 8.078 & 5.45 & 5.48 \\
4999.503 & 0.826 & 0.320 & 238.316 & 8.286 & 4.69 & 4.67 \\
5009.645 & 0.021 & -2.200 & 159.203 & 6.120 & 5.41 & 5.44 \\
5016.161 & 0.848 & -0.480 & 174.672 & 3.573 & 4.95 & 4.93 \\
5024.844 & 0.818 & -0.530 & 176.244 & 5.328 & 4.99 & 4.97 \\
5043.584 & 0.836 & -1.590 & 123.686 & 5.131 & 5.04 & 5.02 \\
5145.460 & 1.460 & -0.540 & 135.898 & 7.219 & 5.20 & 5.15 \\
5147.478 & 0.000 & -1.940 & 181.900 & 11.725 & 5.63 & 5.65 \\
5230.967 & 2.239 & -1.190 & 29.035 & 0.177 & 4.69 & 4.66 \\
5282.376 & 1.053 & -1.810 & 101.973 & 0.890 & 5.04 & 5.02 \\
5300.010 & 1.053 & -2.300 & 75.916 & 0.836 & 4.99 & 4.98 \\
5338.306 & 0.826 & -2.730 & 67.137 & 1.479 & 4.91 & 4.88 \\
5366.639 & 0.818 & -2.460 & 75.554 & 2.440 & 4.78 & 4.74 \\
5384.630 & 0.826 & -2.770 & 57.245 & 0.779 & 4.78 & 4.75 \\
5453.642 & 1.443 & -1.600 & 76.460 & 2.076 & 4.85 & 4.80 \\
5465.773 & 1.067 & -2.910 & 29.690 & 0.205 & 4.76 & 4.75 \\
5471.192 & 1.443 & -1.420 & 88.090 & 0.473 & 4.89 & 4.82 \\
5866.451 & 1.067 & -0.790 & 162.541 & 2.731 & 5.08 & 5.06 \\
5918.536 & 1.067 & -1.640 & 111.474 & 0.589 & 4.94 & 4.92 \\
5922.110 & 1.046 & -1.380 & 127.487 & 1.442 & 4.93 & 4.94 \\
5937.809 & 1.067 & -1.940 & 95.440 & 0.507 & 4.91 & 4.92 \\
6091.171 & 2.267 & -0.320 & 87.629 & 0.475 & 4.92 & 4.84 \\
6336.099 & 1.443 & -1.690 & 77.201 & 0.419 & 4.86 & 4.79 \\
6395.472 & 1.502 & -2.540 & 22.009 & 0.152 & 4.78 & 4.76 \\
6497.684 & 1.443 & -2.020 & 65.279 & 0.363 & 4.97 & 4.92 \\
6554.223 & 1.443 & -1.150 & 117.196 & 0.666 & 4.99 & 4.89 \\
6556.062 & 1.460 & -1.060 & 128.324 & 1.458 & 5.11 & 5.02 \\
8692.329 & 1.046 & -2.130 & 110.286 & 2.062 & 4.94 & 4.86 \\
\noalign{\smallskip}
\multicolumn{7}{c}{ \ion{Ti}{II} }\\
\noalign{\smallskip}
4798.531 & 1.080 & -2.660 & 106.218 & 12.979 & 5.06 & 5.06 \\
4865.610 & 1.116 & -2.700 & 107.891 & 2.043 & 5.18 & 5.18 \\
4874.009 & 3.095 & -0.860 & 57.401 & 0.589 & 4.93 & 4.79 \\
5005.167 & 1.566 & -2.730 & 68.903 & 2.361 & 4.93 & 4.92 \\
5211.530 & 2.590 & -1.410 & 65.531 & 1.450 & 4.93 & 4.92 \\
\noalign{\smallskip}
\hline
\noalign{\smallskip}
\multicolumn{7}{c}{ 122563 }\\
\noalign{\smallskip}
\noalign{\smallskip}
\multicolumn{7}{c}{ \ion{Ti}{I} }\\
\noalign{\smallskip}
3598.714 & 0.900 & -0.610 & 9.777 & 0.220 & 2.35 & 2.74 \\
3725.152 & 1.067 & -0.340 & 4.023 & 0.142 & 1.84 & 2.19 \\
3729.807 & 0.000 & -0.280 & 62.128 & 0.445 & 2.06 & 2.42 \\
3741.059 & 0.021 & -0.150 & 60.851 & 0.428 & 1.91 & 2.30 \\
3752.858 & 0.048 & 0.050 & 74.038 & 0.929 & 2.03 & 2.40 \\
3904.783 & 0.900 & 0.150 & 26.459 & 0.363 & 2.05 & 2.51 \\
3924.526 & 0.021 & -0.870 & 29.245 & 0.373 & 2.03 & 2.40 \\
3947.768 & 0.021 & -0.890 & 31.212 & 0.770 & 2.09 & 2.44 \\
3958.205 & 0.048 & -0.110 & 73.019 & 2.408 & 2.10 & 2.48 \\
3998.636 & 0.048 & 0.020 & 75.584 & 1.503 & 2.01 & 2.40 \\
4008.927 & 0.021 & -1.000 & 26.877 & 0.330 & 2.10 & 2.44 \\
4025.077 & 2.153 & -1.040 & 90.895 & 0.846 & 2.77 & 2.78 \\
4287.403 & 0.836 & -0.370 & 19.025 & 0.231 & 2.25 & 2.48 \\
4449.142 & 1.887 & 0.470 & 5.708 & 0.152 & 2.10 & 2.35 \\
4518.022 & 0.826 & -0.250 & 18.905 & 0.202 & 2.09 & 2.33 \\
4534.776 & 0.836 & 0.350 & 43.721 & 0.333 & 2.00 & 2.25 \\
4535.569 & 0.826 & 0.140 & 36.210 & 0.479 & 2.07 & 2.31 \\
4548.764 & 0.826 & -0.280 & 17.749 & 0.190 & 2.08 & 2.32 \\
4555.483 & 0.848 & -0.400 & 13.090 & 0.158 & 2.08 & 2.32 \\
4617.269 & 1.749 & 0.440 & 8.111 & 0.138 & 2.11 & 2.37 \\
4981.731 & 0.848 & 0.570 & 62.468 & 0.396 & 2.04 & 2.30 \\
5022.868 & 0.826 & -0.330 & 17.917 & 0.195 & 2.09 & 2.34 \\
5036.464 & 1.443 & 0.140 & 12.377 & 0.162 & 2.20 & 2.51 \\
5173.743 & 0.000 & -1.060 & 32.679 & 0.316 & 2.10 & 2.46 \\
5192.969 & 0.021 & -0.950 & 37.243 & 0.333 & 2.09 & 2.46 \\
\noalign{\smallskip}
\multicolumn{7}{c}{ \ion{Ti}{II} }\\
\noalign{\smallskip}
3500.333 & 0.122 & -2.040 & 99.922 & 2.727 & 2.68 & 2.70 \\
3533.854 & 2.061 & -1.310 & 15.915 & 0.250 & 2.30 & 2.34 \\
3535.407 & 2.061 & 0.010 & 66.974 & 1.476 & 2.10 & 2.19 \\
3757.685 & 1.566 & -0.440 & 83.136 & 2.103 & 2.20 & 2.33 \\
3759.291 & 0.607 & 0.280 & 209.168 & 1.279 & 2.62 & 2.62 \\
3761.321 & 0.574 & 0.180 & 194.250 & 4.133 & 2.56 & 2.57 \\
3761.872 & 2.590 & -0.420 & 34.370 & 0.552 & 2.38 & 2.44 \\
3774.647 & 0.574 & -2.650 & 55.363 & 3.764 & 2.61 & 2.63 \\
3776.053 & 1.582 & -1.240 & 51.734 & 0.689 & 2.36 & 2.43 \\
3813.388 & 0.607 & -1.890 & 100.697 & 5.068 & 2.88 & 2.91 \\
3900.539 & 1.130 & -0.290 & 133.115 & 0.853 & 2.74 & 2.72 \\
3987.606 & 0.607 & -2.730 & 48.878 & 0.636 & 2.58 & 2.58 \\
4012.384 & 0.574 & -1.780 & 103.308 & 2.660 & 2.73 & 2.74 \\
4028.338 & 1.892 & -0.920 & 54.402 & 0.386 & 2.42 & 2.48 \\
4053.821 & 1.893 & -1.070 & 41.757 & 0.460 & 2.34 & 2.40 \\
4161.529 & 1.084 & -2.090 & 51.710 & 0.594 & 2.54 & 2.56 \\
4163.644 & 2.590 & -0.130 & 45.062 & 0.480 & 2.28 & 2.35 \\
4316.794 & 2.048 & -1.620 & 18.131 & 0.207 & 2.54 & 2.60 \\
4330.698 & 1.180 & -2.090 & 49.565 & 0.624 & 2.60 & 2.61 \\
4391.026 & 1.231 & -2.300 & 46.454 & 0.470 & 2.80 & 2.80 \\
4395.839 & 1.243 & -1.930 & 51.120 & 0.621 & 2.53 & 2.55 \\
4409.518 & 1.231 & -2.530 & 21.144 & 0.293 & 2.54 & 2.56 \\
4418.331 & 1.237 & -1.990 & 51.096 & 0.348 & 2.58 & 2.60 \\
4444.554 & 1.116 & -2.200 & 47.123 & 0.503 & 2.57 & 2.58 \\
4488.324 & 3.123 & -0.500 & 10.120 & 0.159 & 2.38 & 2.41 \\
4493.522 & 1.080 & -2.780 & 18.654 & 0.202 & 2.53 & 2.55 \\
4518.332 & 1.080 & -2.560 & 32.997 & 0.353 & 2.63 & 2.65 \\
4545.133 & 1.130 & -2.450 & 33.627 & 0.282 & 2.59 & 2.61 \\
4583.409 & 1.165 & -2.840 & 16.976 & 0.221 & 2.64 & 2.65 \\
4609.265 & 1.180 & -3.320 & 5.332 & 0.116 & 2.58 & 2.59 \\
4657.201 & 1.243 & -2.290 & 34.988 & 0.629 & 2.59 & 2.60 \\
4708.663 & 1.237 & -2.350 & 36.021 & 0.280 & 2.65 & 2.66 \\
4719.511 & 1.243 & -3.320 & 6.008 & 0.130 & 2.70 & 2.71 \\
4763.883 & 1.221 & -2.400 & 35.184 & 0.346 & 2.66 & 2.68 \\
4764.525 & 1.237 & -2.690 & 18.544 & 0.198 & 2.60 & 2.62 \\
4798.531 & 1.080 & -2.660 & 28.559 & 0.650 & 2.62 & 2.63 \\
4865.610 & 1.116 & -2.700 & 24.999 & 0.249 & 2.62 & 2.63 \\
4874.009 & 3.095 & -0.860 & 4.867 & 0.123 & 2.35 & 2.40 \\
4911.194 & 3.123 & -0.640 & 8.161 & 0.129 & 2.40 & 2.45 \\
5013.686 & 1.582 & -2.140 & 22.092 & 0.212 & 2.55 & 2.57 \\
5129.156 & 1.892 & -1.340 & 40.354 & 0.823 & 2.48 & 2.50 \\
5185.902 & 1.893 & -1.410 & 35.635 & 0.319 & 2.47 & 2.49 \\
5211.530 & 2.590 & -1.410 & 5.966 & 0.111 & 2.38 & 2.46 \\
5336.786 & 1.582 & -1.600 & 53.625 & 0.370 & 2.56 & 2.58 \\
5381.022 & 1.566 & -1.970 & 34.910 & 0.280 & 2.60 & 2.61 \\
5396.247 & 1.584 & -3.180 & 5.427 & 0.112 & 2.88 & 2.89 \\
5418.768 & 1.582 & -2.130 & 26.487 & 0.208 & 2.61 & 2.62 \\
\noalign{\smallskip}
\hline

%\tablebib{(a)~\citet{Sitnova_2020};
%(b)~\citet{arcturus};
%(c)~\citet{HD84937} and \citet{HD84937-2};
%(d)~\citet{hd14};
%(e)~\citet{anotherhd12}}
\end{longtable}
\end{appendix}


%\begin{thebibliography}{}

%\end{thebibliography}

\end{document}